\documentclass[12pt]{article}
\usepackage{graphicx}
\usepackage{epsfig}
\usepackage{epstopdf}
\DeclareGraphicsExtensions{.pdf,.eps,.png,.jpg,.mps}

\usepackage{cite}

\def\lsim{\mathrel{\rlap{\lower3pt\hbox{\hskip0pt$\sim$}}
     \raise1pt\hbox{$<$}}}         
\def\gsim{\mathrel{\rlap{\lower4pt\hbox{\hskip1pt$\sim$}}
     \raise1pt\hbox{$>$}}}         

\usepackage{amsmath}
\usepackage{amsfonts}

\begin{document}
\begin{titlepage}

\centerline{\Large \bf Heterotic Risk Models}
\medskip

\centerline{Zura Kakushadze$^\S$$^\dag$\footnote{\, Zura Kakushadze, Ph.D., is the President of Quantigic$^\circledR$ Solutions LLC,
and a Full Professor at Free University of Tbilisi. Email: \tt zura@quantigic.com}}
\bigskip

\centerline{\em $^\S$ Quantigic$^\circledR$ Solutions LLC}
\centerline{\em 1127 High Ridge Road \#135, Stamford, CT 06905\,\,\footnote{\, DISCLAIMER: This address is used by the corresponding author for no
purpose other than to indicate his professional affiliation as is customary in
publications. In particular, the contents of this paper
are not intended as an investment, legal, tax or any other such advice,
and in no way represent views of Quantigic$^\circledR$ Solutions LLC,
the website \underline{www.quantigic.com} or any of their other affiliates.
The title of this paper is inspired by the heterotic string theory, with no substantive connection therewith.
}}
\centerline{\em $^\dag$ Free University of Tbilisi, Business School \& School of Physics}
\centerline{\em 240, David Agmashenebeli Alley, Tbilisi, 0159, Georgia}
\medskip
\centerline{(April 30, 2015)}

\bigskip
\medskip

\begin{abstract}
{}We give a complete algorithm and source code for constructing what we refer to as heterotic risk models (for equities), which combine: i) granularity of an industry classification; ii) diagonality of the principal component factor covariance matrix for any sub-cluster of stocks; and iii) dramatic reduction of the factor covariance matrix size in the Russian-doll risk model construction. This appears to prove a powerful approach for constructing out-of-sample stable short-lookback risk models. Thus, for intraday mean-reversion alphas based on overnight returns, Sharpe ratio optimization using our heterotic risk models sizably improves the performance characteristics compared to weighted regressions based on principal components or industry classification. We also give source code for: a) building statistical risk models; and ii) Sharpe ratio optimization with homogeneous linear constraints and position bounds.
\end{abstract}
\medskip
\end{titlepage}

\newpage

\section{Introduction}

{}When the number of stocks in a portfolio is large and the number of available (relevant) observations in the historical time series of returns is limited -- which is essentially a given for short-horizon quantitative trading strategies -- the sample covariance matrix (SCM) is badly singular. This makes portfolio optimization -- {\em e.g.}, Sharpe ratio maximization -- challenging as it requires the covariance matrix to be invertible. A standard method for circumventing this difficulty is to employ factor models,\footnote{\,
For a partial list of literature related to factor risk models, see, {\em e.g.},
\cite{Q1}, \cite{Q2}, \cite{Q3}, \cite{Q4}, \cite{Q5}, \cite{Q6}, \cite{Q7}, \cite{Q8}, \cite{Q9}, \cite{Q10},
\cite{Q11}, \cite{Q12}, \cite{Q13}, \cite{Q14}, \cite{Q15}, \cite{Q16}, \cite{Q17}, \cite{Q18}, \cite{Q19}, \cite{Q20},
\cite{Q21}, \cite{Q22}, \cite{Q23}, \cite{Q24}, \cite{Q26, Q25}, \cite{Q27}, \cite{Q28}, \cite{Q29, Q30},
\cite{Q31, Q32, Q33}, \cite{Q34}, \cite{Q35}, \cite{Q36}, \cite{Q37, Q38, Q39, Q40},
\cite{Q41}, \cite{Q42, Q43}, \cite{Q44, Q45}, \cite{Q46}, \cite{Q47}, \cite{Q48}, \cite{Q49}, \cite{Q50},
\cite{Q51}, \cite{Q52}, \cite{Q53}, \cite{Q54}, \cite{Q55, Q56}, \cite{4F, MeanRev, RusDoll}, \cite{CustomRM},
\cite{Q62}, \cite{Q63}, \cite{Q64}, \cite{Q65}, \cite{Q66}, \cite{Q67}, \cite{Q68}, \cite{Q69}, \cite{Q70},
\cite{Q71}, \cite{Q72}, \cite{Q73}, \cite{Q74, Q75}, \cite{Q76}, \cite{Q77}, \cite{Q78}, \cite{Q79}, \cite{Q80},
\cite{Q81}, \cite{Q82}, \cite{Q83}, \cite{Q84}, \cite{Q85, Q86, Q87}, \cite{Q88}, \cite{Q89}, \cite{Q90, Q91}, \cite{Q92},
\cite{Q93, Q94}, \cite{Q95, Q96}, \cite{Q97}, \cite{Q98}, \cite{Q99}, \cite{Q100}, \cite{Q101}, \cite{Q102},
and references therein.
}
which, instead of computing SCM for a large number of stocks, allow to compute a factor covariance matrix (FCM) for many fewer risk factors. However, the number of relevant risk factors itself can be rather large. {\em E.g.}, in a (desirably) granular industry classification (IC), the number of industries\footnote{\, By this we mean the stock clusters at the most granular level in the IC hierarchy. {\em E.g.}, in BICS these would be sub-industries, whereas other ICs have different naming conventions.} can be in 3 digits for a typical (liquid) trading universe. Then, even the sample FCM can be singular.

{}In (Kakushadze, 2015c) a simple idea is set forth: model FCM itself via a factor model, and repeat this process until the remaining FCM is small enough and can be computed. In fact, at the end of this process we may even end up with a single factor, for which ``FCM" is simply its variance.\footnote{\, Generally, off-diagonal elements of a sample (stock or factor) covariance matrix tend to be unstable out-of-sample, whereas its diagonal elements (variances) typically are much more stable.} This construction -- termed as ``Russian-doll" risk models (Kakushadze, 2015c) -- dramatically reduces the number of or altogether eliminates the factors for which (off-diagonal) FCM must be computed. The ``catch" is that at each successive step we must: i) identify the risk factors; and ii) compute the specific (idiosyncratic) risk (ISR) and FCM consistently.

{}Identifying the risk factors in the Russian-doll construction is facilitated by using a binary industry classification:\footnote{\, The number of non-binary style factors is at most of order 10 and does not pose a difficultly for computing the factor covariance matrix. It is the ubiquitous industry factors that are problematic.} using BICS as an illustrative example, industries serve as the risk factors for sub-industries; sectors -- there are only 10 of them -- serve as the risk factors for industries; and -- if need be -- the ``market" serves as the sole risk factor for sectors. Correctly computing ISR and FCM is more nontrivial: the algorithms for this are generally deemed proprietary. One method in the ``lore" is to use a cross-sectional linear regression, where the returns are regressed over a factor loadings matrix (FLM), and FCM is identified with the (serial) covariance matrix of the regression coefficients, whereas ISR squared is identified with the (serial) variance of the regression residuals. However, as discussed in (Kakushadze, 2015c), generally this does not satisfy a nontrivial requirement (which is often overlooked in practice) that the factor model reproduce the historical in-sample total variance.

{}In this paper we share a complete algorithm and source code for building what we refer to as ``heterotic" risk models. It is based on a simple observation that, if we use principal components (PCs) as FLM, the aforementioned total variance condition is automatically satisfied. Unfortunately, the number of useful PCs is few as it is limited by the number of observations, and they also tend to be unstable out-of-sample (as they are based on off-diagonal covariances), with the first PC being most stable. We circumvent this by building FLM from the first PCs of the blocks (sub-matrices) of the sample correlation matrix\footnote{\, And not SCM -- this is an important technical detail, see the discussion in Section \ref{cor.not.cov}.} corresponding to -- in the BICS language -- the sub-industries. {\em I.e.}, if there are $N$ stocks and $K$ sub-industries, FLM is $N\times K$, and in each column the elements corresponding to the tickers in that sub-industry are proportional to the first PC of the corresponding block, with all other elements vanishing.\footnote{\, Note that this is not the same as a``hybrid" (mixture) of industry and statistical risk factors.} The total variance condition is automatically satisfied. Then, applying the Russian-doll construction yields a nonsingular factor model covariance matrix, which, considering it sizably adds value in Sharpe ratio optimization for certain intraday mean-reversion alphas we backtest, appears to be stable out-of-sample.

{}Heterotic risk models are based on our proprietary know-how. We hope sharing it with the investment community encourages organic custom risk models building.

{}This paper is organized as follows. In Section \ref{sec2} we briefly review some generalities of factor models and discuss in detail the total variance condition. In Section \ref{sub.pc} we discuss the PC approach and an algorithm for fixing the number of PC factors, with the R source code in Appendix \ref{app.A}. We discuss heterotic risk models in detail in Section \ref{sec.het}, with the complete Russian-doll embedding in Section \ref{sec.RD} and the R source code in Appendix \ref{app.B}. In Section \ref{sec.horse} we run a horse race of intraday mean-reversion alphas via i) weighted regressions and ii) optimization using heterotic risk models. For optimization with homogeneous linear constrains and (liquidity/position) bounds we use the R source code in Appendix \ref{app.C}.\footnote{\, The source code given in the appendices is not written to be ``fancy" or optimized for speed or in any other way. Its sole purpose is to illustrate the algorithms described in the main text in a simple-to-understand fashion. Some legalese relating to this code is given in Appendix \ref{app.D}.} We briefly conclude in Section \ref{sec.conc}.

\newpage

\section{Multi-factor Risk Models}\label{sec2}
\subsection{Generalities}

{}In a multi-factor risk model, a sample covariance matrix $C_{ij}$ for $N$ stocks, $i,j = 1,\dots,N$, which is computed based on a time series of stock returns $R_i$ ({\em e.g.}, daily close-to-close returns), is modeled via a constructed covariance matrix $\Gamma_{ij}$ given by
\begin{eqnarray}\label{Gamma}
 &&\Gamma \equiv \Xi + \Omega~\Phi~\Omega^T\\
 && \Xi_{ij} \equiv \xi_i^2 ~\delta_{ij}
\end{eqnarray}
where $\delta_{ij}$ is the Kronecker delta; $\Gamma_{ij}$ is an $N\times N$ matrix; $\xi_i$ is the specific risk (a.k.a. idiosyncratic risk) for each stock; $\Omega_{iA}$ is an $N\times K$ factor loadings matrix; and $\Phi_{AB}$ is a $K\times K$ factor covariance matrix, $A,B=1,\dots,K$, where $K\ll N$. {\em I.e.}, the random processes $\Upsilon_i$ corresponding to $N$ stock returns are modeled via $N$ random processes $\chi_i$ (specific risk) together with $K$ random processes $f_A$ (factor risk):
\begin{eqnarray}\label{Upsilon}
 &&\Upsilon_i = \chi_i + \sum_{A=1}^K \Omega_{iA}~f_A\\
 &&\mbox{Cov}(\chi_i, \chi_j) = \Xi_{ij}\\
 &&\mbox{Cov}(\chi_i, f_A) = 0\\
 &&\mbox{Cov}(f_A, f_B) = \Phi_{AB}\\
 &&\mbox{Cov}(\Upsilon_i, \Upsilon_j) = \Gamma_{ij}
\end{eqnarray}
When $M<N$, where $M+1$ is the number of observations in each time series, the sample covariance matrix $C_{ij}$ is singular with $M$ nonzero eigenvalues. In contrast, assuming all $\xi_i>0$ and $\Phi_{AB}$ is positive-definite, then $\Gamma_{ij}$ is automatically positive-definite (and invertible). Furthermore, the off-diagonal elements of $C_{ij}$ typically are not expected to be too stable out-of-sample. On the contrary, the factor model covariance matrix $\Gamma_{ij}$ is expected to be much more stable as the number of risk factors, for which the factor covariance matrix $\Phi_{AB}$ needs to be computed, is $K\ll N$.

\subsection{Conditions on Total Variances}

{}The prime aim of a risk model is to predict the covariance matrix out-of-sample as precisely as possible, including the out-of-sample total variances. However, albeit this requirement is often overlooked in practical applications, a well-built factor model had better reproduce the in-sample total variances. That is, we require that the factor model total variance $\Gamma_{ii}$ coincide with the in-sample total variance $C_{ii}$:
\begin{equation}\label{tot.risk}
 \Gamma_{ii} = \xi_i^2 + \sum_{A,B=1}^K \Omega_{iA}~\Phi_{AB}~\Omega_{iB} = C_{ii}
\end{equation}
{\em A priori} this gives $N$ conditions\footnote{\, With additional assumptions not all of these conditions are nontrivial (see below).} for $N + K(K+1)/2$ unknowns $\xi_i$ and $\Phi_{AB}$, so we need additional assumptions\footnote{\, There are no ``natural" $K(K+1)/2$ conditions we can impose on $\Gamma_{ij}$, $i\neq j$ in terms of out-of-sample unstable $C_{ij}$, $i\neq j$. Note that the variances $C_{ii}$ typically are much more stable.} to compute $\xi_i$ and $\Phi_{AB}$.

\subsection{Linear Regression}

{}One such assumption -- {\em intuitively} -- is that the total risk should be attributed to the factor risk to the greatest extent possible, {\em i.e.}, the part of the total risk attributed to the specific risk should be minimized. One way to formulate this requirement mathematically is via least squares. First, mimicking (\ref{Upsilon}), we decompose the stock returns $R_i$ via a linear model
\begin{equation}
 R_i = \epsilon_i + \sum_{A=1}^K \Omega_{iA}~f_A
\end{equation}
Here the residuals $\epsilon_i$ are {\em not} the same as $\chi_i$ in (\ref{Upsilon}); in particular, generally the covariance matrix $\mbox{Cov}(\epsilon_i, \epsilon_j)$ is not diagonal (see below). We can require that
\begin{equation}\label{lin.w}
 \sum_{i=1}^N z_i~\epsilon_i^2 \rightarrow \mbox{min}
\end{equation}
where $z_i > 0$, and the minimization is w.r.t. $f_A$. This produces a weighted linear regression\footnote{\, Without the intercept, that is, unless the intercept is already subsumed in $\Omega_{iA}$.} with the regression weights $z_i$. So, what should these weights be?

\subsection{Correlations, Not Covariances}\label{cor.not.cov}

{}While choosing unit weights $z_i\equiv 1$ might appear as the simplest thing to do, this suffers from a shortcoming. Intuitively it is clear that -- {\em on average} -- the residuals $\epsilon_i$ are larger for more volatile stocks, so the regression with unit weights would produce skewed results.\footnote{\, Cross-sectionally, stock volatility typically has a roughly log-normal distribution.} This can be readily rectified using nontrivial regression weights. A ``natural" choice is $z_i = 1/C_{ii}$. In fact, we have a regression with unit weights:
\begin{eqnarray}
 &&{\widetilde R}_i = {\widetilde \epsilon}_i + \sum_{A=1}^K {\widetilde \Omega}_{iA}~f_A\\
 &&\sum_{i=1}^N {\widetilde \epsilon}_i^2 \rightarrow \mbox{min}\label{lin.unit}
\end{eqnarray}
where ${\widetilde R}_i \equiv R_i /\sqrt{C_{ii}}$, ${\widetilde \Omega}_{iA} \equiv \Omega_{iA} /\sqrt{C_{ii}}$, and
${\widetilde \epsilon}_i \equiv \epsilon_i /\sqrt{C_{ii}}$ on average are expected to be much more evenly distributed compared with $\epsilon_i$ -- we have scaled away the volatility skewness via rescaling the returns, factor loadings and residuals by $\sqrt{C_{ii}}$.

{}So, we are now modeling the sample {\em correlation} matrix $\Psi_{ij} \equiv C_{ij}/ \sqrt{C_{ii}}\sqrt{C_{jj}}$ (note that $\Psi_{ii}=1$, while $|\Psi_{ij}|\leq 1$ for $i\neq j$)\footnote{\, If $R_i(t_s)$ ($t_s$ labels the observations in the time series and in the above notations the index $s$ takes $M+1$ values) are the time series of the stock returns based on which the sample covariance matrix $C_{ij}$ is computed (so $C_{ii} = \mbox{Var}(R_i(t_s))$, where the variance is serial), then $\Psi_{ij}$ is the sample covariance matrix for the ``normalized" returns ${\widetilde R}_i(t_s) \equiv R_i(t_s)/\sqrt{C_{ii}}$, {\em i.e.}, $\Psi_{ij} = \mbox{Cov}({\widetilde R}_i(t_s), {\widetilde R}_j(t_s)) = \mbox{Cor}(R_i(t_s), R_j(t_s))$, where $\mbox{Cov}(\cdot, \cdot)$ and $\mbox{Cor}(\cdot, \cdot)$ are serial.} via another factor model matrix
\begin{equation}
 {\widetilde\Gamma}_{ij} = {\widetilde \xi}^2_i~\delta_{ij} + \sum_{A,B=1}^K {\widetilde \Omega}_{iA}~\Phi_{AB}~{\widetilde \Omega}_{jB}
\end{equation}
where ${\widetilde\Gamma}_{ij} \equiv \Gamma_{ij}/\sqrt{C_{ii}}\sqrt{C_{jj}}$, and ${\widetilde \xi}^2_i\equiv \xi_i^2/C_{ii}$. The solution to (\ref{lin.unit}) is given by (in matrix notation)
\begin{eqnarray}\label{fac.ret}
 &&f = \left({\widetilde \Omega}^T~{\widetilde\Omega}\right)^{-1}{\widetilde\Omega}^T~{\widetilde R}\\
 &&{\widetilde \epsilon} = \left[1 - Q\right] {\widetilde R}\\
 &&Q \equiv {\widetilde \Omega}\left({\widetilde \Omega}^T~{\widetilde\Omega}\right)^{-1}{\widetilde\Omega}^T
\end{eqnarray}
where $Q$ is a projection operator: $Q^2 = Q$. Consequently, we have:
\begin{eqnarray}
 &&{\widehat \Xi} \equiv \mbox{Cov}\left({\widetilde\epsilon},{\widetilde \epsilon}^T\right) = \left[1 - Q\right] \Psi \left[1 - Q^T\right]\\
 &&{\widetilde \Omega}~\Phi~{\widetilde \Omega}^T = {\widetilde \Omega}~\mbox{Cov}\left(f, f^T\right){\widetilde \Omega}^T = Q~\Psi~Q^T
\end{eqnarray}
Note that the matrix ${\widehat \Xi}$ is not diagonal. However, the idea here is to identify ${\widetilde\xi}_i^2$ with the diagonal part of ${\widehat \Xi}$:
\begin{equation}\label{xi}
 {\widetilde \xi}_i^2 \equiv {\widehat \Xi}_{ii} = \left(\left[1 - Q\right] \Psi \left[1 - Q^T\right]\right)_{ii}
\end{equation}
and we have
\begin{equation}
 {\widetilde \Gamma}_{ij} = {\widetilde \xi}_i^2~\delta_{ij} + \left(Q~\Psi~Q^T\right)_{ij}
\end{equation}
Note that ${\widetilde \xi}_i^2$ defined via (\ref{xi}) are automatically positive (nonnegative, to be precise -- see below). However, we must satisfy the conditions (\ref{tot.risk}), which reduce to
\begin{equation}
 {\widetilde\Gamma}_{ii} = {\widetilde\xi}_i^2 + \sum_{A,B=1}^K {\widetilde\Omega}_{iA}~\Phi_{AB}~{\widetilde\Omega}_{iB} = \Psi_{ii} = 1
\end{equation}
and imply
\begin{eqnarray}\label{T}
 &&T_{ii} = 0\\
 &&T \equiv 2~Q~\Psi~Q^T - Q~\Psi - \Psi~Q^T
\end{eqnarray}
The $N$ conditions (\ref{T}) are not all independent. Thus, we have $\mbox{Tr}(T) = 0$.

\section{Principal Components}\label{sub.pc}

{}The conditions (\ref{T}) are nontrivial. They are not satisfied for an arbitrary factor loadings matrix ${\widetilde \Omega}_{iA}$. However, there is a simple way of satisfying these conditions, to wit, by building ${\widetilde \Omega}_{iA}$ from the principal components of the correlation matrix $\Psi_{ij}$.

{}Let $V_i^{(a)}$, $a=1,\dots,N$ be the $N$ principal components of $\Psi_{ij}$ forming an orthonormal basis
\begin{eqnarray}
 &&\sum_{j=1}^N \Psi_{ij}~V_j^{(a)} = \lambda^{(a)}~V_i^{(a)}\\
 &&\sum_{i=1}^N V_i^{(a)}~V_i^{(b)} = \delta_{ab}
\end{eqnarray}
such that the eigenvalues $\lambda^{(a)}$ are ordered decreasingly: $\lambda^{(1)} > \lambda^{(2)} >\dots$. More precisely, some eigenvalues may be degenerate. For simplicity -- and this is not critical here -- we will assume that all positive eigenvalues are non-degenerate. However, we can have multiple null eigenvalues. Typically, the number of nonvanishing eigenvalues\footnote{\, This number can be smaller if some stock returns are 100\% correlated or anti-correlated. For the sake of simplicity -- and this not critical here -- we will assume that there are no such returns.} is $M$, where, as above, $M+1$ is the number of observations in the stock return time series. We can readily construct a factor model with $K\leq M$:
\begin{equation}\label{FLM.PC}
 {\widetilde \Omega}_{iA} = \sqrt{\lambda^{(A)}}~V_i^{(A)}
\end{equation}
Then the factor covariance matrix $\Phi_{AB} = \delta_{AB}$ and we have
\begin{eqnarray}\label{PC}
 && {\widetilde \Gamma}_{ij} = {\widetilde \xi}_i^2~\delta_{ij} + \sum_{A=1}^K \lambda^{(A)}~V_i^{(A)}~V_j^{(A)}\\
 &&{\widetilde \xi}_i^2 = 1 - \sum_{A=1}^K \lambda^{(A)}\left(V_i^{(A)}\right)^2
\end{eqnarray}
so ${\widetilde \Gamma}_{ii} = \Psi_{ii} = 1$. See Appendix \ref{app.B} for the R code including the following algorithm.

\subsection{Fixing $K$}\label{sub.fix.K}

{}When $K = M$ we have ${\widetilde \Gamma} = \Psi$, which is singular.\footnote{\, Note that ${\widetilde \xi}_i^2 = \sum_{a=K+1}^M \lambda^{(a)}\left(V_i^{(a)}\right)^2 \geq 0$. We are assuming $\lambda^{(a)}\geq 0$, which (up to computational precision) is the case if there are no N/As in the stock return times series.} Therefore, we must have $K \leq K_{max} < M$. So, how do we determine $K_{max}$? And is there $K_{min}$ other than the evident answer $K_{min} = 1$? Here we can do a lot of complicated, even convoluted things. Or we can take a pragmatic approach and come up with a simple heuristic. Here is one simple algorithm that does a very decent job at fixing $K$.

{}The idea here is simple. It is based on the observation that, as $K$ approaches $M$, $\mbox{min}({\widetilde \xi}^2_i)$ goes to 0 ({\em i.e.}, less and less of the total risk is attributed to the specific risk, and more and more of it is attributed to the factor risk), while as $K$ approaches 0, $\mbox{max}({\widetilde \xi}^2_i)$ goes to 1 ({\em i.e.}, less and less of the total risk is attributed to the factor risk, and more and more of it is attributed to the specific risk). So, as a rough cut, we can think of $K_{max}$ and $K_{min}$ as the maximum/minimum values of $K$ such that $\mbox{min}({\widetilde \xi}^2_i) \geq \zeta^2_{min}$ and $\mbox{max}({\widetilde \xi}^2_i)\leq \zeta^2_{max}$, where $\zeta_{min}$ and $\zeta_{max}$ are some desired bounds on the fraction of the contribution of the specific risk into the total risk. {\em E.g.}, we can set $\zeta_{min} = 10\%$ and $\zeta_{max}=90\%$. In practice, we actually need to fix the value of $K$, not $K_{max}$ and $K_{min}$, especially that for some preset values of $\zeta_{min}$ and $\zeta_{max}$ we may end up with $K_{max} < K_{min}$. However, the above discussion aids us in coming up with a simple heuristic definition for what $K$ should be. Here is one:
\begin{eqnarray}\label{K}
 &&|g(K) - 1| \rightarrow \mbox{min}\\
 \label{g}
 &&g(K) \equiv \sqrt{\mbox{min}({\widetilde \xi}^2_i)} + \sqrt{\mbox{max}({\widetilde \xi}^2_i)}
\end{eqnarray}
{\em i.e.}, we take $K$ for which $g(K)$ (which monotonically decreases with increasing $K$) is closest to 1. This simple algorithm works pretty well in practical applications.\footnote{\, The distribution of ${\widetilde \xi}^2_i$ is skewed; typically, ${\widetilde \xi}^2_i$ has a tail at higher values, while $\ln({\widetilde \xi}^2_i)$ has a tail at lower values, and the distribution is only roughly log-normal. So $K$ is not (the floor/cap of) $M/2$, but somewhat higher, albeit close to it. See Table \ref{table.prin.comp} and Figure 1 for an illustrative example.}

\subsection{Limitations}

{}An evident limitation of the principal component approach is that the number of risk factors is limited by $M$. If long lookbacks are unavailabe/undesirable, as, {\em e.g.}, in short-holding quantitative trading strategies, then typically $M \ll N$. Yet, the number of the actually relevant underlying risk factors can be substantially greater than $M$, and most of these risk factors are missed by the principal component approach. In this regard, we can ask: can we use other than the first $M$ principal components to build a factor model? The answer, prosaically, is that, without some additional information, it is unclear what to do with the principal components with null eigenvalues. They simply do not contribute to any sample factor covariance matrix. However, not all is lost. There is a way around this difficulty.

\section{Heterotic Construction}\label{sec.het}
\subsection{Industry Risk Factors}

{}Without long lookbacks, the number of risk factors based on principal components is limited.\footnote{\, The number of style factors is also limited (especially for short horizons), of order 10 or fewer.} However, risk factors based on a granular enough industry classification can be plentiful. Furthermore, they are independent of the pricing data and, in this regard, are insensitive to the lookback. In fact, typically they tend to be rather stable out-of-sample as companies seldom jump industries, let alone sectors.

{}For terminological definiteness, here we will use the BICS nomenclature for the levels in the industry classification, albeit this is not critical here. Also, BICS has three levels ``sector $\rightarrow$ industry $\rightarrow$ sub-industry" (where ``sub-industry" is the most granular level). The number of levels in the industry hierarchy is not critical here either. So, we have: $N$ stocks labeled by $i=1,\dots,N$; $K$ sub-industries labeled by $A=1,\dots,K$; $F$ industries labeled by $a=1,\dots,F$; and $L$ sectors labeled by $\alpha=1,\dots,L$. More generally, we can think of such groupings as ``clusters". Sometimes, loosely, we will refer to such cluster based factors as ``industry" factors.\footnote{\, Albeit in the BICS context we may be referring to, {\em e.g.}, sub-industries, while in other classification schemes the actual naming may be altogether different.}

\subsection{``Binary" Property}

{}The binary property implies that each stock belongs to one and only one sub-industry, industry and sector (or, more generally, cluster). Let $G$ be the map between stocks and sub-industries, $S$ be the map between sub-industries and industries, and $W$ be the map between industries and sectors:
\begin{eqnarray}\label{G.map}
 &&G:\{1,\dots,N\}\mapsto\{1,\dots,K\}\\
 &&S:\{1,\dots,K\}\mapsto\{1,\dots,F\}\label{S.map}\\
 &&W:\{1,\dots,F\}\mapsto\{1,\dots,L\}\label{W.map}
\end{eqnarray}
The nice thing about the binary property is that the clusters (sub-industries, industries and sectors) can be used to identify blocks (sub-matrices) in the correlation matrix $\Psi_{ij}$. {\em E.g.}, at the most granular level, for sub-industries, the binary matrix $\delta_{G(i), A}$ defines such blocks. Thus, the sum $B_A\equiv \sum_{i=1}^N \delta_{G(i), A}~X_i$, where $X_i$ is an arbitrary $N$-vector, is the same as $\sum_{i\in J(A)} X_i$, where $J(A)$ is the set of tickers in the sub-industry $A$. These blocks are the backbone of the following construction.

\subsection{Heterotic Models}

{}Consider the following factor loadings matrix:
\begin{eqnarray}\label{ind.pc}
 &&{\widetilde \Omega}_{iA} = \delta_{G(i), A}~U_i\\
 &&U_i \equiv [U(A)]_i,~~~i\in J(A),~~~A=1,\dots K
\end{eqnarray}
where $J(A)\equiv \{i|G(i)=A\}$ is the set of tickers (whose number $N(A)\equiv |J(A)|$) in the sub-industry labeled by $A$. Then the $N(A)$-vector $U(A)$ is the {\em first} principal component of the $N(A) \times N(A)$ matrix $\Psi(A)$ defined via $[\Psi(A)]_{ij} \equiv \Psi_{ij}$, $i,j\in J(A)$. (Note that $\sum_{i\in J(A)} [U(A)]_i^2 =1$; also, let the corresponding (largest) eigenvalue of $\Psi(A)$ be $\lambda(A)$.)\footnote{\, If $N(A)=1$, {\em i.e.}, we have only one ticker in the sub-industry labeled by $A$, then $[U(A)]_i =1$ and $\lambda(A) = \Psi_{ii} = 1$, $i\in J(A)$.} With this factor loadings matrix we can compute the factor covariance matrix and specific risk via a linear regression as above, and we get:
\begin{eqnarray}
 &&{\widetilde \xi}_i^2 = 1 - \lambda(G(i))~ U_i^2\\
 &&\left({\widetilde \Omega}~\Phi~{\widetilde \Omega}^T\right)_{ij} = U_i~U_j \sum_{k\in J(G(i))} ~\sum_{l\in J(G(j))} U_k ~\Psi_{kl}~ U_l
\end{eqnarray}
so we have\footnote{\, For single-ticker sub-industries ($N(A) = 1$) the specific risk vanishes: ${\widetilde\xi}^2_i=0$; however, this does not pose a problem as this does not cause the matrix ${\widetilde\Gamma}_{ij}$ to be singular (see below).}
\begin{equation}
 {\widetilde \Gamma}_{ij} = \left[1 - \lambda(G(i))~ U_i^2\right]\delta_{ij} + U_i ~U_j \sum_{k\in J(G(i))}~ \sum_{l\in J(G(j))} U_k ~\Psi_{kl}~ U_l
\end{equation}
and automatically ${\widetilde \Gamma}_{ii} = 1$. This simplicity is due to the use of the (first) principal components corresponding to the blocks $\Psi(A)$ of the sample correlation matrix.

\subsubsection{Multiple Principal Components}

{}For the sake of completeness, let us discuss an evident generalization. Above in (\ref{ind.pc}) we took the binary map between the tickers and sub-industries and augmented it with the first principal components of the corresponding blocks in the sample correlation matrix. Instead of taking only the first principal component, we can take the first $P(A)\geq 1$ principal components for each block labeled by the sub-industry $A$ ($A=1,\dots, K$). Then we have ${\widehat K} = \sum_{A=1}^K P(A)$ risk factors labeled by pairs ${\widehat A} \equiv (A, I)$, where for a given value of $A$ we have $I\in D(A)$ (with $|D(A)| = P(A)$). The factor loadings matrix reads:
\begin{equation}
 {\widetilde \Omega}_{i{\widehat A}} = \delta_{G(i), A}~[U(A)]_i^{(I)}
\end{equation}
where $U(A)$ is the $N(A)\times P(A)$ matrix whose columns are the first $P(A)$ principal components (with eigenvalues $[\lambda(A)]^{(I)}$) of the $N(A) \times N(A)$ matrix $\Psi(A)$ (as above, $[\Psi(A)]_{ij} \equiv \Psi_{ij}$, $i,j\in J(A)$, and $\sum_{i\in J(A)} [U(A)]_i^{(I)}~[U(A)]_i^{(J)} = \delta_{IJ}$, $I,J\in D(A)$.) In order to have nonvanishing specific risks, it is necessary that we take $P(A) < M$ ($M+1$ is the number of observations in the time series). We then have
\begin{eqnarray}
 &&\Gamma_{ij} = \left[1 - \sum_{I\in D(G(i))} [\lambda(G(i))]^{(I)}~ \left([U(G(i))]^{(I)}_i\right)^2\right]\delta_{ij} +\nonumber\\
 &&\,\,\,\,\,\,\,+\sum_{I\in D(G(i))}~\sum_{J\in D(G(j))} [U(G(i))]_i^{(I)}~[U(G(j))]_j^{(J)}\times\nonumber\\
 &&\,\,\,\,\,\,\,\times\sum_{k\in J(G(i))}~ \sum_{l\in J(G(j))} [U(G(i))]_k^{(I)} ~\Psi_{kl}~ [U(G(j))]_l^{(J)}
\end{eqnarray}
and (as in the case above with all $P(A)\equiv 1$) automatically ${\widetilde \Gamma}_{ii} = 1$.

\subsubsection{Caveats}

{}The above construction might look like a free lunch, but it is not. Let us start with the $P(A)\equiv 1$ case (first principal components only). For short lookbacks, the number of risk factors typically is too large: $K$ can easily be greater than $M$, so\footnote{\, Note that $\Phi_{AA} = \lambda(A)$.}
\begin{equation}
 \Phi_{AB} = \sum_{i\in J(A)}~\sum_{j\in J(B)} U_i~\Psi_{ij}~U_j
\end{equation}
is singular. In general, the sample factor covariance matrix is singular if $K > M$. We will deal with this issue below via the nested Russian-doll risk model construction.

{}This issue is further exacerbated in the multiple principal component construction (with at least some $P(A)>1$) as the number of risk factors ${\widehat K} > K$ is even larger. This too can be dealt with via the Russian-doll construction. However, there is yet another caveat pertinent to using multiple principal components, irrespective of whether the factor covariance matrix is singular or not. The principal components are based on off-diagonal elements of $\Psi_{ij}$ and tend to be unstable out-of-sample, the first principal component typically being the most stable. So, for the sake of simplicity, below we will focus on the case with only first principal components.

\section{Russian-Doll Construction}\label{sec.RD}
\subsection{General Idea}

{}As discussed above, the sample factor covariance matrix $\Phi_{AB}$ is singular if the number of factors $K$ is greater than $M$. The simple idea behind the Russian-doll construction is to model such $\Phi_{AB}$ itself via yet another factor model matrix $\Gamma^\prime_{AB}$ (as opposed to computing it as a sample covariance matrix of the risk factors $f_A$):\footnote{\, We use a prime on $\Gamma^\prime_{AB}$, ${\widetilde \xi}^{\prime}_A$, $\Phi_{ab}^\prime$, {\em etc.} to avoid confusion with $\Gamma_{ij}$, ${\widetilde \xi}_i$, $\Phi_{AB}$, {\em etc.}}
\begin{eqnarray}
 &&\Gamma^\prime_{AB} = \sqrt{\Phi_{AA}}\sqrt{\Phi_{BB}}~{\widetilde \Gamma}^\prime_{AB}\\
 &&{\widetilde \Gamma}^\prime_{AB} = ({\widetilde \xi}^{\prime}_A)^2~\delta_{AB} + \sum_{a,b=1}^F {\widetilde\Omega}^\prime_{Aa}~\Phi^\prime_{ab}~{\widetilde\Omega}^\prime_{Bb}
\end{eqnarray}
where ${\widetilde \xi}^\prime_A$ is the specific risk for the ``normalized" factor return ${\widetilde f}_A \equiv f_A/\sqrt{\Phi_{AA}}$; ${\widetilde \Omega}^\prime_{Aa}$, $A=1,\dots,K$, $a=1,\dots,F$ is the corresponding factor loadings matrix; and $\Phi^\prime_{ab}$ is the factor covariance matrix for the underlying risk factors $f^\prime_a$, $a=1,\dots,F$, where we assume that $F\ll K$. If the smaller factor covariance matrix $\Phi^\prime_{ab}$ is still singular, we model it via yet another factor model with fewer risk factors, and so on -- until the resulting factor covariance matrix is nonsingular. If, at the final stage, we are left with a single factor, then the resulting $1\times 1$ factor covariance matrix is automatically nonsingular -- it is simply the sample variance of the remaining factor.

\subsection{Complete Heterotic Russian-Doll Embedding}\label{sub.het.rd}

{}For concreteness we will use the BICS terminology for the levels in the industry classification, albeit this is not critical here. Also, BICS has three levels ``sector $\rightarrow$ industry $\rightarrow$ sub-industry" (where ``sub-industry" is the most granular level). For definiteness, we will assume three levels here, and the generalization to more levels is straightforward. So, we have: $N$ stocks labeled by $i=1,\dots,N$; $K$ sub-industries labeled by $A=1,\dots,K$; $F$ industries labeled by $a=1,\dots,F$; and $L$ sectors labeled by $\alpha=1,\dots,L$. A nested Russian-doll risk model then is constructed as follows:
\begin{eqnarray}\label{Gamma.RD}
 &&\Gamma_{ij} = \sqrt{C_{ii}}\sqrt{C_{jj}}~{\widetilde \Gamma}_{ij}\\
 &&{\widetilde \Gamma}_{ij} = {\widetilde \xi}_i^2~\delta_{ij} + U_i~U_j~\Gamma^\prime_{G(i), G(j)}\\
 &&\Gamma^\prime_{AB} = \sqrt{\Phi_{AA}}\sqrt{\Phi_{BB}}~{\widetilde\Gamma}^\prime_{AB}\label{Gamma.prime.RD}\\
 &&{\widetilde\Gamma}^\prime_{AB} = ({\widetilde \xi}^\prime_A)^2~\delta_{AB} + U^\prime_A~U^\prime_B~\Gamma^{\prime\prime}_{S(A), S(B)}\\
 &&\Gamma^{\prime\prime}_{ab} = \sqrt{\Phi^\prime_{aa}}\sqrt{\Phi^\prime_{bb}}~{\widetilde\Gamma}^{\prime\prime}_{ab}\\
 &&{\widetilde\Gamma}^{\prime\prime}_{ab} = ({\widetilde \xi}^{\prime\prime}_a)^2~\delta_{ab} + U^{\prime\prime}_a~U^{\prime\prime}_b~\Gamma^{\prime\prime\prime}_{W(a), W(b)}\\
 &&\Gamma^{\prime\prime\prime}_{\alpha\beta} = \sqrt{\Phi^{\prime\prime}_{\alpha\alpha}}\sqrt{\Phi^{\prime\prime}_{\beta\beta}}~{\widetilde \Gamma}^{\prime\prime\prime}_{\alpha\beta}\\
 &&{\widetilde \Gamma}^{\prime\prime\prime}_{\alpha\beta} = ({\widetilde \xi}^{\prime\prime\prime}_\alpha)^2~\delta_{\alpha\beta} +  U^{\prime\prime\prime}_\alpha~U^{\prime\prime\prime}_\beta~\Phi^{\prime\prime\prime}
\end{eqnarray}
where
\begin{eqnarray}\label{xi.RD}
 &&{\widetilde \xi}_i^2 = 1 - \lambda(G(i))~U_i^2\\
 &&({\widetilde \xi}^\prime_A)^2 = 1 - \lambda^\prime(S(A))~(U^\prime_A)^2\\
 &&({\widetilde \xi}^{\prime\prime}_a)^2 = 1 - \lambda^{\prime\prime}(W(a))~(U^{\prime\prime}_a)^2\\
 &&({\widetilde \xi}^{\prime\prime\prime}_\alpha)^2 = 1 - \lambda^{\prime\prime\prime}~(U^{\prime\prime\prime}_\alpha)^2
\end{eqnarray}
and
\begin{eqnarray}
 &&\Phi_{AB} = \sum_{i\in J(A)}~\sum_{j\in J(B)} U_i~\Psi_{ij}~U_j\\
 &&\Phi^\prime_{ab} = \sum_{A\in J^\prime(a)} ~\sum_{B\in J^\prime(b)} U^\prime_A~\Psi^\prime_{AB}~U^\prime_B\\
 &&\Phi^{\prime\prime}_{\alpha\beta} = \sum_{a\in J^{\prime\prime}(\alpha)} ~\sum_{b\in J^{\prime\prime}(\beta)} U^{\prime\prime}_a~ \Psi^{\prime\prime}_{ab}~U^{\prime\prime}_b\label{Theta.1}\\
 &&\Phi^{\prime\prime\prime}= \sum_{\alpha,\beta = 1}^L U^{\prime\prime\prime}_\alpha~\Psi^{\prime\prime\prime}_{\alpha\beta}~U^{\prime\prime\prime}_\beta
\end{eqnarray}
so we have $\Phi_{AA} = \lambda(A)$, $\Phi^\prime_{aa} = \lambda^\prime(a)$, $\Phi^{\prime\prime}_{\alpha\alpha} = \lambda^{\prime\prime}(\alpha)$, and $\Phi^{\prime\prime\prime} = \lambda^{\prime\prime\prime}$ (see below). Also, $J(A) = \{i| G(i)=A\}$ ($N_A \equiv |J(A)|$ tickers in sub-industry $A$), $J^\prime(a) = \{A|S(A) = a\}$ ($N^\prime(a) \equiv |J^\prime(a)|$ sub-industries in industry $a$), $J^{\prime\prime}(\alpha) = \{a|W(a) = \alpha\}$ ($N^{\prime\prime}(\alpha)\equiv |J^{\prime\prime}(\alpha)|$ industries in sector $\alpha$), and the maps $G$ (tickers to sub-industries), $S$ (sub-industries to industries) and $W$ (industries to sectors) are defined in (\ref{G.map}), (\ref{S.map}) and (\ref{W.map}). Furthermore,
\begin{eqnarray}
 &&\Psi_{ij} = C_{ij} / \sqrt{C_{ii}}\sqrt{C_{jj}}\\
 &&\Psi^\prime_{AB} = \Phi_{AB} / \sqrt{\Phi_{AA}}\sqrt{\Phi_{BB}}\\
 &&\Psi^{\prime\prime}_{ab} = \Phi^\prime_{ab} / \sqrt{\Phi^\prime_{aa}}\sqrt{\Phi^\prime_{bb}}\\
 &&\Psi^{\prime\prime\prime}_{\alpha\beta} = \Phi^{\prime\prime}_{\alpha\beta} / \sqrt{\Phi^{\prime\prime}_{\alpha\alpha}}\sqrt{\Phi^{\prime\prime}_{\beta\beta}}
\end{eqnarray}
and
\begin{eqnarray}\label{U.RD}
 &&U_i \equiv [U(A)]_i,~~~i\in J(A)\\
 &&U^\prime_A \equiv [U^\prime(a)]_A,~~~A\in J^\prime(a)\\
 &&U^{\prime\prime}_a \equiv [U^{\prime\prime}(\alpha)]_a,~~~a\in J^{\prime\prime}(\alpha)
\end{eqnarray}
The $N(A)$-vector $U(A)$ is the first principal component of $\Psi(A)$ with the eigenvalue $\lambda(A)$ ($[\Psi(A)]_{ij} \equiv \Psi_{ij}$, $i,j\in J(A)$), the $N^\prime(a)$-vector $U^\prime(a)$ is the first principal component of $\Psi^\prime(a)$ with the eigenvalue $\lambda^\prime(a)$ ($[\Psi^\prime(a)]_{AB} \equiv \Psi^\prime_{AB}$, $A,B\in J^\prime(a)$), the $N^{\prime\prime}(\alpha)$-vector $U^{\prime\prime}(\alpha)$ is the first principal component of $\Psi^{\prime\prime}(\alpha)$ with the eigenvalue $\lambda^{\prime\prime}(\alpha)$ ($[\Psi^{\prime\prime}(\alpha)]_{ab} \equiv \Psi^{\prime\prime}_{ab}$, $a,b\in J^{\prime\prime}(\alpha)$), while $U^{\prime\prime\prime}_\alpha$ is the first principal component of $\Psi^{\prime\prime\prime}_{\alpha\beta}$ with the eigenvalue $\lambda^{\prime\prime\prime}$. The vectors $U(A)$, $U^\prime(a)$ and $U^{\prime\prime}(\alpha)$ are normalized, so $\sum_{i\in J(A)} U_i^2 = 1$, $\sum_{A\in J^\prime(a)} (U^\prime_A)^2 = 1$, $\sum_{a\in J^{\prime\prime}(\alpha)} (U^{\prime\prime}_a)^2 = 1$, and also $\sum_{\alpha=1}^L (U^{\prime\prime\prime}_\alpha)^2 = 1$.

{}For the sake of completeness, above we included the step where the sample factor covariance matrix $\Phi^{\prime\prime}_{\alpha\beta}$ for the sectors is further approximated via a 1-factor model ${\widetilde \Gamma}^{\prime\prime\prime}_{\alpha\beta}$. If $\Phi^{\prime\prime}_{\alpha\beta}$ computed via (\ref{Theta.1}) is nonsingular, then this last step can be omitted,\footnote{\, That is, assuming there are enough observations in the time series for out-of-sample stability.} so at the last stage we have $L$ factors (as opposed to a single factor).\footnote{\, This last factor can be interpreted as the ``market" risk factor. For the sake of completeness, the definitions of the factors at each stage are as follows: (i) for the sub-industries $f_A = \sum_{i\in J(A)} U_i~{\widetilde R}_i$, where ${\widetilde R}_i = R_i/\sqrt{C_{ii}}$; (ii) for the industries $f^\prime_a = \sum_{A\in J^\prime(a)} U^\prime_A~{\widetilde f}_A$, where ${\widetilde f}_A = f_A/\sqrt{\Phi_{AA}}$; (iii) for the sectors $f^{\prime\prime}_\alpha = \sum_{a\in J^{\prime\prime}(\alpha)} U^{\prime\prime}_a~{\widetilde f}^\prime_a$, where ${\widetilde f}^\prime_a = f^\prime_a/\sqrt{\Phi^\prime_{aa}}$; and (iv) for the ``market" $f^{\prime\prime\prime} = \sum_{\alpha=1}^L U^{\prime\prime\prime}_\alpha~{\widetilde f}^{\prime\prime}_\alpha$, where ${\widetilde f}^{\prime\prime}_\alpha = f^{\prime\prime}_\alpha/\sqrt{\Phi^{\prime\prime}_{\alpha\alpha}}$.} Similarly, if we have enough observations to compute the sample covariance matrix $\Phi^\prime_{ab}$ for the industries, we can stop at that stage. Finally, note that in the above construction we are guaranteed to have $({\widetilde \xi}^{\prime\prime\prime}_\alpha)^2 > 0$, $({\widetilde \xi}^{\prime\prime}_a)^2 >0$, $({\widetilde \xi}^\prime_A)^2> 0$ and ${\widetilde \xi}_i^2\geq 0$ (with the last equality occurring only for single-ticker sub-industries and not posing a problem -- see below).\footnote{\, For a typical, large trading universe, industries and sectors usually contain more than one ticker; however, there can be cases of single-ticker sub-industries. Nonetheless, ${\widetilde\Gamma}_{ij}$ is nonsingular. Indeed, for an arbitrary $N$-vector $X_i$ we have $X^T~{\widetilde \Gamma}~X > 0$ unless $X_i \equiv 0$, $i\not\in H$, where $H\equiv\{i|N(G(i))=1\}$. For such $X_i$ we have $X^T~{\widetilde \Gamma}~X = \sum_{A,B\in E} Y_A~\Gamma^\prime_{AB}~Y_B > 0$, where $E\equiv\{A|N(A)=1\}$, $Y_A \equiv\delta_{A, G(i)} X_i$, $A\in E$, and we have taken into account that by construction $\Gamma^\prime_{AB}$ (and its sub-matrix with $A,B\in E$) is positive-definite, and also that $U_i = 1$, $i\in H$. More on this below.} In Appendix \ref{app.B} we give the R code for building heterotic risk models.

\subsection{Model Covariance Matrix and Its Inverse}\label{sub.inv}

{}The model covariance matrix is given by $\Gamma_{ij}$ defined in (\ref{Gamma.RD}). For completeness, let us present it in the ``canonical" form:
\begin{equation}
 \Gamma_{ij} = \xi_i^2~\delta_{ij} + \sum_{A,B=1}^K \Omega_{iA}~\Phi^*_{AB}~\Omega_{jB}
\end{equation}
where
\begin{eqnarray}
 &&\xi_i^2 \equiv C_{ii}~{\widetilde \xi}_i^2\\
 &&\Omega_{iA} \equiv \sqrt{C_{ii}}~U_i~\delta_{G(i),A}\\
 &&\Phi^*_{AB} \equiv \Gamma^\prime_{AB}
\end{eqnarray}
where ${\widetilde \xi}_i^2$ is defined in (\ref{xi.RD}), $U_i$ is defined in (\ref{U.RD}), $\Gamma^\prime_{AB}$ is defined in (\ref{Gamma.prime.RD}), and we use the star superscript in the our factor covariance matrix $\Phi^*_{AB}$ (which is nonsingular) to distinguish it from the sample factor covariance matrix $\Phi_{AB}$ (which is singular).

{}In many applications, such as portfolio optimization, one needs the inverse of the matrix $\Gamma$. When we have no single-ticker sub-industries, the inverse is given by (in matrix notation)
\begin{eqnarray}\label{Gamma.inv}
 &&\Gamma^{-1} = \Xi^{-1} - \Xi^{-1}~\Omega~\Delta^{-1}~\Omega^T~\Xi^{-1}\\
 &&\Delta \equiv (\Phi^*)^{-1} + \Omega^T~\Xi^{-1}~\Omega\\
 &&\Xi\equiv \mbox{diag}(\xi^2_i)
\end{eqnarray}
However, when there are some single-ticker sub-industries, the corresponding $\xi_i^2=0$, $i\in H$ ($H\equiv \{i|N(G(i)) = 1\}$), so (\ref{Gamma.inv}) ``breaks". Happily, there is an easy ``fix". This is because for such tickers the specific risk and factor risk are {\em indistinguishable}. Recall that $U_i = 1$, $i\in H$, and $\Phi^*_{AA} = 1$, $A\in E$ ($E\equiv\{A|N(A)=1\}$). We can rewrite $\Gamma_{ij}$ via
\begin{equation}
 \Gamma_{ij} = {\widehat \xi}_i^2~\delta_{ij} + \sum_{A,B=1}^K \Omega_{iA}~{\widehat \Phi}^*_{AB}~\Omega_{jB}
\end{equation}
where: ${\widehat \xi}_i^2 = \xi_i^2$ for $i\not\in H$; ${\widehat \xi}_i^2 = C_{ii}~\zeta_i$ for $i\in H$ with arbitrary $\zeta_i$, $0 < \zeta_i \leq 1$; ${\widehat \Phi}^*_{AB} = \Phi^*_{AB}$ if $A\not\in E$ or $B\not\in E$ or $A\neq B$; and ${\widehat \Phi}^*_{AA} = 1 - \delta_{A, G(i)}~\zeta_i$ for $A\in E$. (Here we have taken into account that $U_i=1$, $i\in H$.) Now we can invert $\Gamma$ via
\begin{eqnarray}\label{Gamma.inv.1}
 &&\Gamma^{-1} = {\widehat \Xi}^{-1} - {\widehat \Xi}^{-1}~\Omega~{\widehat \Delta}^{-1}~\Omega^T~{\widehat \Xi}^{-1}\\
 &&{\widehat \Delta} \equiv ({\widehat \Phi}^*)^{-1} + \Omega^T~{\widehat \Xi}^{-1}~\Omega\\
 &&{\widehat \Xi}\equiv \mbox{diag}({\widehat \xi}^2_i)
\end{eqnarray}
Note that, due to the factor model structure, to invert the $N\times N$ matrix $\Gamma$, we only need to invert two $K\times K$ matrices ${\widehat \Phi}^*$ and ${\widehat \Delta}$. If there are no single-ticker sub-industries, then $\Phi^*$ itself has a factor model structure and involves inverting two $F\times F$ matrices, one of which has a factor model structure, and so on.

\section{Horse Race}\label{sec.horse}

{}So, suppose we have built a complete heterotic risk model. How do we know it adds value? {\em I.e.}, how do we know that the off-diagonal elements of the factor model covariance matrix $\Gamma_{ij}$ are stable out-of-sample to the extent that they add value. We can run a horse race. There are many ways of doing this. Here is one. For a given trading universe we compute some expected returns, {\em e.g.}, based on overnight mean-reversion. We can construct a trading portfolio by using our heterotic risk model covariance matrix in the optimization whereby we maximize the Sharpe ratio (subject to the dollar neutrality constraint). On the other hand, we can run the same optimization with a diagonal sample covariance matrix $\mbox{diag}(C_{ii})$ subject to neutrality (via linear homogeneous constraints) w.r.t. the underlying heterotic risk factors (plus dollar neutrality).\footnote{\, For comparative purposes, we will also run separate backtests where we require neutrality w.r.t. BICS sub-industries and principal components.} In fact, optimization with such diagonal covariance matrix and subject to such linear homogeneous constraints is equivalent to a weighted cross-sectional regression with the loadings matrix (over which the expected returns are regressed) identified with the factor loadings matrix (augmented by the intercept, {\em i.e.}, the unit vector, for dollar neutrality) and the regression weights identified with the inverse sample variances $1/C_{ii}$ (see (Kakushadze, 2015a) for details). So, we will refer to the horse race as between optimization (using the heterotic risk model) and weighted regression (with the aforementioned linear homogeneous constraints).\footnote{\, The remainder of this section somewhat overlaps with Section 7 of (Kakushadze, 2015a) as the backtesting models are similar, albeit not identical.}

\subsection{Notations}

{}Let $P_{is}$ be the time series of stock prices, where $i=1,\dots,N$ labels the stocks, and $s=0,1,\dots,M$ labels the trading dates, with $s=0$ corresponding to the most recent date in the time series. The superscripts $O$ and $C$ (unadjusted open and close prices) and $AO$ and $AC$ (open and close prices fully adjusted for splits and dividends) will distinguish the corresponding prices, so, {\em e.g.}, $P^C_{is}$ is the unadjusted close price. $V_{is}$ is the unadjusted daily volume (in shares). Also, for each date $s$ we define the overnight return as the previous-close-to-open return:
\begin{equation}\label{c2o.ret}
 E_{is} \equiv \ln\left({P^{AO}_{is} / P^{AC}_{i,s+1}}\right)
\end{equation}
This return will be used in the definition of the expected return in our mean-reversion alpha. We will also need the close-to-close return
\begin{equation}\label{c2c.ret}
 R_{is} \equiv \ln\left({P^{AC}_{is} / P^{AC}_{i,s+1}}\right)
\end{equation}
An out-of-sample (see below) time series of these returns will be used in constructing the heterotic risk model and computing, among other things, the sample variances $C_{ii}$. Also note that all prices in the definitions of $E_{is}$ and $R_{is}$ are fully adjusted.

{}We assume that: i) the portfolio is established at the open\footnote{\, This is a so-called ``delay-0" alpha: the same price, $P^O_{is}$ (or adjusted $P^{AO}_{is}$), is used in computing the expected return (via $E_{is}$) and as the establishing fill price.} with fills at the open prices $P^O_{is}$; ii) it is liquidated at the close on the same day -- so this is a purely intraday alpha -- with fills at the close prices $P^C_{is}$; and iii) there are no transaction costs or slippage -- our aim here is not to build a realistic trading strategy, but to test that our heterotic risk model adds value to the alpha. The P\&L for each stock
\begin{equation}
 \Pi_{is} = H_{is}\left[{P^C_{is}\over P^O_{is}}-1\right]
\end{equation}
where $H_{is}$ are the {\em dollar} holdings. The shares bought plus sold (establishing plus liquidating trades) for each stock on each day are computed via $Q_{is} = 2 |H_{is}| / P^O_{is}$.

\subsection{Universe Selection}

{}For the sake of simplicity,\footnote{\, In practical applications, the trading universe of liquid stocks typically is selected based on market cap, liquidity (ADDV), price and other (proprietary) criteria.} we select our universe based on the average daily dollar volume (ADDV) defined via (note that $A_{is}$ is out-of-sample for each date $s$):
\begin{equation}\label{ADDV}
 A_{is}\equiv {1\over d} \sum_{r=1}^d V_{i, s+r}~P^C_{i, s+r}
\end{equation}
We take $d=21$ ({\em i.e.}, one month), and then take our universe to be the top 2000 tickers by ADDV. To ensure that we do not inadvertently introduce a universe selection bias, we rebalance monthly (every 21 trading days, to be precise). {\em I.e.}, we break our 5-year backtest period (see below) into 21-day intervals, we compute the universe using ADDV (which, in turn, is computed based on the 21-day period immediately preceding such interval), and use this universe during the entire such interval. We do have the survivorship bias as we take the data for the universe of tickers as of 9/6/2014 that have historical pricing data on http://finance.yahoo.com (accessed on 9/6/2014) for the period 8/1/2008 through 9/5/2014. We restrict this universe to include only U.S. listed common stocks and class shares (no OTCs, preferred shares, {\em etc.}) with BICS sector, industry and sub-industry assignments as of 9/6/2014.\footnote{\, The choice of the backtesting window is based on what data was readily available.} However, as discussed in detail in Section 7 of (Kakushadze, 2015a), the survivorship bias is not a leading effect in such backtests.\footnote{\, Here we are after the {\em relative outperformance}, and it is reasonable to assume that, to the leading order, individual performances are affected by the survivorship bias approximately equally.}

\subsection{Backtesting}

{}We run our simulations over a period of 5 years (more precisely, 1260 trading days going back from 9/5/2014, inclusive). The annualized return-on-capital (ROC) is computed as the average daily P\&L divided by the intraday investment level $I$ (with no leverage) and multiplied by 252. The annualized Sharpe Ratio (SR) is computed as the daily Sharpe ratio multiplied by $\sqrt{252}$. Cents-per-share (CPS) is computed as the total P\&L divided by the total shares traded.\footnote{\, As mentioned above, we assume no transaction costs, which are expected to reduce the ROC of the optimization and weighted regression alphas by the same amount as the two strategies trade the exact same amount by design. Therefore, including the transaction costs would have no effect on the actual {\em relative outperformance} in the horse race, which is what we are after here.}

\subsection{Weighted Regression Alphas}\label{sub.reg}

{}We will always require that our portfolio be dollar neutral:
\begin{equation}\label{DN}
 \sum_{i=1}^N H_{is}= 0
\end{equation}
We will further require neutrality
\begin{equation}
 \sum_{i=1}^N H_{is}~\Lambda_{iAs}= 0
\end{equation}
with the three different incarnations for the loadings matrix $\Lambda_{iA}$ (for each trading day $s$, so we omit the index $s$)\footnote{\, The loadings $\Lambda_{iA}$ in (\ref{load.pc}) and (\ref{load.het}) are computed for each trading date $s$ (as opposed to, say, every 21 days -- see below); in (\ref{load.sub}) they change only with the universe (every 21 days).\label{fn.load}} defined via:
\begin{eqnarray}\label{load.pc}
 &&\mbox{principal components:}~~~\Lambda_{iA} = \sqrt{C_{ii}}~\sqrt{\lambda^{(A)}}~V^{(A)}_i,~~~A=1,\dots,K_{PC}\\
 &&\mbox{sub-industries:}~~~\Lambda_{iA} = \delta_{G(i),A},~~~A=1,\dots,K\label{load.sub}\\
 &&\mbox{heterotic risk factors:}~~~\Lambda_{iA} = \sqrt{C_{ii}}~U_i~\delta_{G(i),A},~~~A=1,\dots,K\label{load.het}
\end{eqnarray}
Here $V^{(A)}_i$ are the first $K_{PC}$ principal components (with the eigenvalues $\lambda^{(A)}$)\footnote{\, The factor $\sqrt{\lambda^{(A)}}$ in (\ref{load.pc}) does not affect the regression residuals below.} of the sample correlation matrix $\Psi_{ij}$. For each date $s$ we take $M+1=d=21$ trading days as our lookback ({\em i.e.}, the number of observations) in the out-of-sample time series of close-to-close (see (\ref{c2c.ret})) returns $(R_{i,(s+1)}, R_{i,(s+2)}, \dots, R_{i,(s+d)})$ (based on which we compute the sample covariance (correlation) matrix $C_{ijs}$ ($\Psi_{ijs}$) for each $s$), so the number of the nonvanishing eigenvalues $\lambda^{(A)}>0$ is $M=20$, and we take $K_{PC}=M$. Further, the map $G$ between tickers and sub-industries is defined in (\ref{G.map}), and $K$ is the number of sub-industries.\footnote{\, In (\ref{load.sub}) we deliberately take $\Lambda_{iA} = \delta_{G(i),A}$ as opposed to $\Lambda_{iA} = \sqrt{C_{ii}}~\delta_{G(i),A}$ (see below). Note that with (\ref{load.sub}) the intercept is subsumed in $\Lambda_{iA}$ as we have $\sum_{A=1}^K \Lambda_{iA} = 1$, so (\ref{DN}) is automatic.} Finally, the vector $U_i$ in (\ref{load.het}) is defined in (\ref{U.RD}).

{}For each date labeled by $s$, we run cross-sectional regressions of the overnight (see (\ref{c2o.ret})) returns $E_{is}$ over the corresponding loadings matrix, call it $Y$ (with indices suppressed), which has 3 different incarnations: i) for principal components, $Y$ is an $N\times (K_{PC}+1)$ matrix, whose first column in the intercept (unit $N$-vector), and the remaining columns are populated by $\Lambda_{iA}$ defined in (\ref{load.pc}); ii) for sub-industries, the elements of $Y$ are the same as $\Lambda_{iA}$ defined in (\ref{load.sub}); and iii) for heterotic risk factors, $Y$ is an $N\times (K+1)$ matrix, whose first column in the intercept (unit $N$-vector), and the remaining columns are populated by $\Lambda_{iA}$ defined in (\ref{load.het}). We take the regression weights to be $z_i \equiv 1/C_{ii}$. More precisely, to avoid unnecessary variations in the weights $z_i$ (as such variations could result in unnecessary overtrading), we do not recompute $z_i$ daily but every 21 trading days, same as with the trading universe.

{}In the cases i)-iii) above, we compute the residuals $\varepsilon_{is}$ of the weighted regression and the dollar holdings $H_{is}$ via (we use matrix notation and suppress indices):
\begin{eqnarray}
 &&{\widetilde E} \equiv Z~\varepsilon = Z \left[E - Y~(Y^T~Z~Y)^{-1}~Y^T~Z~E\right]\\
 &&H_{is} \equiv -{\widetilde E}_{is} ~ {I\over\sum_{j=1}^N \left|{\widetilde E}_{js}\right|}
\end{eqnarray}
where $Z\equiv\mbox{diag}(z_i)$, we have dollar neutrality (\ref{DN}),\footnote{\, Due to ${\widetilde E}_{is}$ having 0 cross-sectional means, which in turn is due to the intercept either being included (the cases i) and iii)), or being subsumed in the loadings matrix $Y$ (the case ii)).} and $\sum_{i=1}^N \left|H_{is}\right| = I$ (the total {\em intraday} dollar investment level (long plus short), which is the same for all dates $s$).

{}The simulation results are given in Table \ref{table2} and P\&Ls for the 3 cases i)-iii) are plotted in Figure 2. For comparison purposes -- and to alley any potential concerns that the results in Table \ref{table2} may not hold for realistic position bounds, in Table \ref{table3} we give the simulation results for the same cases i)-iii) above with the strict bounds
\begin{equation}\label{liq}
 |H_{is}| \leq 0.01~A_{is}
\end{equation}
so not more than 1\% of each stock's ADDV is bought or sold. We use the bounded regression algorithm and the R source code of (Kakushadze, 2015b) to run these simulations. Expectedly, the liquidity bounds (\ref{liq}) lower ROC and CPS while improving SR, but in the same fashion for all 3 weighted regression alphas i)-iii). The results in Tables \ref{table2} and \ref{table3} confirm our prior intuitive argument that the sub-industries outperform the principal components simply because they are more numerous.\footnote{\, In Table \ref{table2} the heterotic risk factors outperform the sub-industries. However, this is largely an artifact of defining $\Lambda_{iA}$ as in (\ref{load.sub}). If we take $\Lambda_{iA} = \sqrt{C_{ii}}~\delta_{G(i),A}$ instead (and augment the regression loadings matrix $Y$ with the intercept for dollar neutrality), we will get (without the bounds (\ref{liq}) -- the results with the bounds are similar): ROC = 51.62\%, SR = 13.45, CPS = 2.26.}

{}If we compute $\Lambda_{iA}$ in (\ref{load.pc}) and (\ref{load.het}) every 21 trading days (instead of daily -- see fn. \ref{fn.load}), the difference is very slight. {\em E.g.}, for the heterotic risk factors computed every 21 days (with no bounds) we get: ROC = 51.66\%, SR = 13.42, CPS = 2.26.

{}Finally, let us also mention that, in the weighted regressions ii) and iii), the dollar holdings for the tickers in the single-ticker sub-industries are automatically null. This is {\em not} the case for optimized alphas (see below). Generally, if single-ticker (or small) sub-industries are undesirable, one
can ``prune" the industry hierarchy tree by merging (single-ticker and/or small) sub-industries at the industry level.

\newpage
\subsection{Optimized Alphas}\label{sub.opt}

{}As mentioned above, our goal is to determine whether the heterotic risk model construction adds value by comparing the simulated performance of the weighted regression alphas above to the simulated performance of the optimized alphas (via maximizing the Sharpe ratio) based on the same expected returns $E_{is}$. In maximizing the Sharpe ratio, we use the heterotic risk model covariance matrix $\Gamma_{ij}$ given by (\ref{Gamma.RD}), which we compute every 21 trading days (same as for the universe). For each date (we omit the index $s$) we maximize the Sharpe ratio subject to the dollar neutrality constraint:
\begin{eqnarray}
 &&{\cal S} \equiv {\sum_{i=1}^N H_i~E_i\over{\sqrt{\sum_{i,j=1}^N \Gamma_{ij}~H_i~H_j}}} \rightarrow \mbox{max}\\
 &&\sum_{i=1}^N H_i = 0\label{d.n.opt}
\end{eqnarray}
The solution is given by
\begin{equation}\label{H.opt}
 H_i = -\gamma \left[\sum_{j = 1}^N \Gamma^{-1}_{ij}~E_j - \sum_{j=1}^N \Gamma^{-1}_{ij}~{{\sum_{k,l=1}^N \Gamma^{-1}_{kl}~E_l}\over{\sum_{k,l = 1}^N \Gamma^{-1}_{kl}}}\right]
\end{equation}
where $\Gamma^{-1}$ is the inverse of $\Gamma$ (see Subsection \ref{sub.inv}), and the overall normalization constant $\gamma > 0$ (this is a mean-reversion alpha) is fixed via the requirement that
\begin{equation}
 \sum_{i=1}^N \left|H_i\right| = I
\end{equation}
Note that (\ref{H.opt}) satisfies the dollar neutrality constraint (\ref{d.n.opt}).

{}The simulation results are given in Table \ref{table2} in the bottom row. The P\&L plot for this optimized alpha is included in Figure 2. For the same reasons as in the case of weighted regression alphas, in the bottom row of Table \ref{table3} we give the simulation results for the same optimized alpha with the strict liquidity bounds (\ref{liq}).\footnote{\, In Tables \ref{table2} and \ref{table3} at the final stage the heterotic risk factors are the (10) BICS sectors: there are enough (20) observations in the time series. The 1-factor model gives almost the same results.} We use the optimization algorithm for maximizing the Sharpe ratio subject to linear homogeneous constraints and bounds discussed in (Kakushadze, 2015a).\footnote{\, The source code for this algorithm is not included in (Kakushadze, 2015a), so we include it in Appendix \ref{app.C}. It is similar to the source code of (Kakushadze, 2015b) for the bounded regression.} Also, in the second rows in Tables \ref{table2} and \ref{table3} we have included the simulation results for the optimized alpha where in the optimization we use the risk factor model covariance matrix $\Gamma_{ij}$ based on the principal components discussed in Section \ref{sub.pc}.\footnote{\, This matrix is given by $\Gamma_{ij} =\sqrt{C_{ii}}\sqrt{C_{jj}}~{\widetilde\Gamma}_{ij}$, where ${\widetilde\Gamma}_{ij}$ is defined in (\ref{PC}), and $K$ is determined via the algorithm of Section \ref{sub.fix.K}. For the $d=21$ trading day lookback in our backtests, the value of $K$ fixed by this algorithm turns out to be $K=13$. } From our simulation results in Tables \ref{table2} and \ref{table3} it is evident that the heterotic risk model predicts off-diagonal elements of the covariance matrix (that is, correlations) out-of-sample rather well. Indeed, using it in the optimization sizably improves ROC, SR and CPS compared with the weighted regressions with all three loadings i)-iii) above.\footnote{\, There exist further (proprietary) performance improvements using the heterotic risk model.}

\section{Concluding Remarks}\label{sec.conc}

{}The heterotic risk model construction we discuss in this paper is based on a ``heterosis" of: i) granularity of an industry classification; ii) diagonality of the principal component factor covariance matrix for any sub-cluster of stocks; and iii) dramatic reduction of the size of the factor covariance matrix in the Russian-doll construction. This is a powerful approach, as is evident from the horse race we ran above.

{}Naturally, one may wonder if we can extend our construction to risk models which do not include any statistical risk factors ({\em i.e.}, principal components) or include other non-binary factors such as style factors. A key simplifying feature in the heterotic construction is that the industry classification, which is used as the backbone (and is augmented with the principal components to satisfy the conditions (\ref{tot.risk})), is binary. Once non-binary risk factors are included, it is more nontrivial to compute the specific risk and the factor covariance matrix (such that (\ref{tot.risk}) are satisfied). However, there exist proprietary algorithms for dealing with this, which are outside of the scope of this paper. We hope to make these algorithms a public knowledge elsewhere.

{}One final remark concerns purely statistical risk models based on principal components. Albeit their market share is rather limited, it is unclear why a portfolio manager would be willing to pay for such models considering that they are straightforward to build in-house, especially now that we have provided the source code for constructing them. One argument is that using option implied volatility (which is available only for optionable stocks) to model stock volatility should work better,\footnote{\, In this context, the paper \cite{ImpliedVol} sometimes is referred to.} and if a portfolio manager does not possess the implied volatility data or the know-how for incorporating it into a statistical risk model, he or she would be better off simply buying one from a provider. However, this argument appears to be thin, at best. Nowadays, with ever-shortening lookbacks, it is unclear if the implied volatility indeed adds any value when the risk model is used in actual portfolio optimization for actual alphas. In this regard a new study would appear to be warranted. In any event, as we saw above, heterotic risk models outperform principal component risk models by a significant margin, so one can build heterotic risk models in-house (instead of buying less powerful statistical models) now that this know-how is in the public domain. The only data needed to construct a heterotic risk model is: i) adjusted close prices; and ii) a granular enough binary industry classification, such as GICS, BICS, ICB, {\em etc.} Most quantitative traders already have this data in-house. So, we hope this paper further encourages/aids organic custom risk model building.

\appendix
\section{R Code: Principal Component Risk Model}\label{app.A}

{}In this appendix we give the R (R Package for Statistical Computing, http://www.r-project.org) source code for building a purely statistical risk model (principal components) based on the algorithm we discuss in Section \ref{sub.pc}, including the algorithm for fixing the number of factors $K$ in Section \ref{sub.fix.K}. The code below is essentially self-explanatory and straightforward. It consists of a single function {\tt{\small qrm.cov.pc(ret, use.cor = T)}}. The input is: i) {\tt{\small ret}}, an $N\times d$ matrix of returns ({\em e.g.}, daily close-to-close returns), where $N$ is the number of tickers, $d$ is the number of observations in the time series ({\em e.g.}, the number of trading days), and the ordering of the dates is immaterial; and ii) {\tt{\small use.cor}}, where for {\tt{\small TRUE}} (default) the risk factors are computed based on the principal components of the sample correlation matrix $\Psi_{ij}$, whereas for {\tt{\small FALSE}} they are computed based on the sample covariance matrix $C_{ij}$. The output is a list: {\tt{\small result\$spec.risk}} is the specific risk $\xi_i$ (not the specific variance $\xi_i^2$),
{\tt{\small result\$fac.load}} is the factor loadings matrix $\Omega_{iA} = \sqrt{C_{ii}}~{\widetilde \Omega}_{iA}$, {\tt{\small result\$fac.cov}} is the factor covariance matrix $\Phi_{AB}$ (with the normalization (\ref{FLM.PC}) for the factor loadings matrix, $\Phi_{AB} = \delta_{AB}$), {\tt{\small result\$cov.mat}} is the factor model covariance matrix $\Gamma_{ij} = \sqrt{C_{ii}}\sqrt{C_{jj}}~{\widetilde \Gamma}_{ij}$, and {\tt{\small result\$inv.cov}} is the matrix $\Gamma^{-1}_{ij}$ inverse to $\Gamma_{ij}$.\\
\\
{\tt{\small
\noindent qrm.cov.pc <- function (ret, use.cor = T)\\
\{\\
\indent print("Running qrm.cov.pc()...")\\
\\
\indent tr <- apply(ret, 1, sd)\\
\indent if(use.cor)\\
\indent \indent ret <- ret / tr\\
\\
\indent d <- ncol(ret)\\
\indent x <- t(ret)\\
\indent x <- var(x, x)\\
\indent tv <- diag(x)\\
\indent x <- eigen(x)\\
\\
\indent g.prev <- 999\\
\indent for(k in 1:(d-1))\\
\indent \{\\
\indent \indent u <- x\$values[1:k]\\
\indent \indent v <- x\$vectors[, 1:k]\\
\indent \indent v <- t(sqrt(u) * t(v))\\
\indent \indent x.f <- v \%*\% t(v)\\
\indent \indent x.s <- tv - diag(x.f)\\
\indent \indent z <- x.s / tv\\
\indent \indent g <- abs(sqrt(min(z)) + sqrt(max(z)) - 1)\\
\\
\indent \indent if(g > g.prev)\\
\indent \indent \indent break\\
\\
\indent \indent g.prev <- g\\
\\
\indent \indent spec.risk <- sqrt(x.s)\\
\indent \indent fac.load <- v\\
\indent \indent fac.cov <- diag(1, k)\\
\indent \indent cov.mat <- diag(x.s) + x.f\\
\indent \}\\
\\
\indent k <- k - 1\\
\indent y.s <- 1 / spec.risk / spec.risk\\
\indent v <- fac.load\\
\indent v1 <- y.s * v\\
\indent inv.cov <- diag(y.s) - v1 \%*\% solve(diag(1, k) + t(v) \%*\% v1) \%*\% t(v1)\\
\\
\indent if(use.cor)\\
\indent \{\\
\indent \indent spec.risk <- tr * spec.risk\\
\indent \indent fac.load <- tr * fac.load\\
\indent \indent cov.mat <- tr * t(tr * cov.mat)\\
\indent \indent inv.cov <- t(inv.cov / tr) / tr\\
\indent \}\\
\\
\indent result <- new.env()\\
\indent result\$spec.risk <- spec.risk\\
\indent result\$fac.load <- fac.load\\
\indent result\$fac.cov <- fac.cov\\
\indent result\$cov.mat <- cov.mat\\
\indent result\$inv.cov <- inv.cov\\
\indent result <- as.list(result)\\
\indent return(result)\\
\}
}}

\section{R Code: Heterotic Risk Model}\label{app.B}
{}In this appendix we give the R source code for building the heterotic risk model based on the algorithm we discuss in Section \ref{sub.het.rd}. The code below is essentially self-explanatory and straightforward as it simply follows the formulas in Section \ref{sub.het.rd}. It consists of a single function {\tt{\small qrm.het(ret, ind, mkt.fac = F, rm.sing.tkr = F)}}. The input is: i) {\tt{\small ret}}, an $N\times d$ matrix of returns ({\em e.g.}, daily close-to-close returns), where $N$ is the number of tickers, $d$ is the number of observations in the time series ({\em e.g.}, the number of trading days), and the ordering of the dates is immaterial; ii) {\tt{\small ind}} is a list whose length {\em a priori} is arbitrary, and its elements are populated by the binary matrices (with rows corresponding to tickers, so {\tt{\small dim(ind[[$\cdot$]])[1]}} is $N$) corresponding to the levels in the input binary industry classification hierarchy in the order of decreasing granularity (so, in the BICS case {\tt{\small ind[[1]]}} is the $N\times K$ matrix $\delta_{G(i), A}$ (sub-industries), {\tt{\small ind[[2]]}} is the $N\times F$ matrix $\delta_{G^\prime(i), a}$ (industries), and {\tt{\small ind[[3]]}} is the $N\times L$ matrix $\delta_{G^{\prime\prime}(i), \alpha}$ (sectors), where the map $G$ is defined in (\ref{G.map}) (tickers to sub-industries), $G^\prime \equiv GS$ (tickers to industries), and $G^{\prime\prime} \equiv GSW$ (tickers to sectors), with the map $S$ (sub-industries to industries) defined in (\ref{S.map}), and the map $W$ (industries to sectors) defined in (\ref{W.map})); iii) {\tt{\small mkt.fac}}, where for {\tt{\small TRUE}} at the final step we have a single factor (``market"), while for {\tt{\small FALSE}} (default) the factors correspond to the least granular level in the industry classification hierarchy; and iv) {\tt{\small rm.sing.tkr}}, where for {\tt{\small TRUE}} the tickers corresponding to the single-ticker clusters at the most granular level in the industry classification hierarchy (in the BICS case this would be the sub-industry level) are dropped altogether, while for {\tt{\small FALSE}} (default) the output universe is the same as the input universe. The output is a list: {\tt{\small result\$spec.risk}} is the specific risk $\xi_i$ (not the specific variance $\xi_i^2$),
{\tt{\small result\$fac.load}} is the factor loadings matrix $\Omega_{iA} = \sqrt{C_{ii}}~{\widetilde \Omega}_{iA}$, {\tt{\small result\$fac.cov}} is the factor covariance matrix $\Phi_{AB}$, {\tt{\small result\$cov.mat}} is the factor model covariance matrix $\Gamma_{ij} = \sqrt{C_{ii}}\sqrt{C_{jj}}~{\widetilde \Gamma}_{ij}$, and {\tt{\small result\$inv.cov}} is the matrix $\Gamma^{-1}_{ij}$ inverse to $\Gamma_{ij}$.\\
\\
{\tt{\small
\noindent qrm.het <- function (ret, ind, mkt.fac = F, rm.sing.tkr = F)\\
\{\\
\indent print("Running qrm.het()...")\\
\\
\indent if(rm.sing.tkr)\\
\indent \{\\
\indent \indent bad <- colSums(ind[[1]]) == 1\\
\indent \indent ind[[1]] <- ind[[1]][, !bad]\\
\indent \indent bad <- rowSums(ind[[1]]) == 0\\
\indent \indent for(lvl in 1:length(ind))\\
\indent \indent \indent ind[[lvl]] <- ind[[lvl]][!bad, ]\\
\indent \indent ret <- ret[!bad, ]\\
\indent \}\\
\\
\indent cov.mat <- list()\\
\indent u <- list()\\
\indent flm <- ind\\
\\
\indent calc.load <- function(load, load1)\\
\indent \{\\
\indent \indent x <- colSums(load1)\\
\indent \indent load <- (t(load1) \%*\% load) / x\\
\indent \indent return(load)\\
\indent \}\\
\\
\indent calc.cor.mat <- function(cov.mat)\\
\indent \{\\
\indent \indent tr <- sqrt(diag(cov.mat))\\
\indent \indent cor.mat <- t(cov.mat / tr) / tr\\
\indent \indent return(cor.mat)\\
\indent \}\\
\indent \\
\indent calc.cov.mat <- function(cor.mat, tr)\\
\indent \{\\
\indent \indent cov.mat <- t(cor.mat * tr) * tr\\
\indent \indent return(cov.mat)\\
\indent \}\\
\indent \indent\\
\indent cov.mat[[1]] <- var(t(ret), t(ret))\\
\indent cor.mat <- calc.cor.mat(cov.mat[[1]])\\
\\
\indent for(lvl in 1:length(ind))\\
\indent \{\\
\indent \indent if(lvl > 1)\\
\indent \indent \indent flm[[lvl]] <- calc.load(ind[[lvl]], ind[[lvl-1]]) \\
\\
\indent \indent u[[lvl]] <- rep(NA, nrow(flm[[lvl]]))\\
\\
\indent \indent for(a in 1:ncol(flm[[lvl]]))\\
\indent \indent \{\\
\indent \indent \indent take <- as.logical(flm[[lvl]][, a])\\
\indent \indent \indent x <- cor.mat[take, take]\\
\indent \indent \indent y <- eigen(x)\$vectors\\
\indent \indent \indent y <- y[, 1]\\
\indent \indent \indent u[[lvl]][take] <- y\\
\indent \indent \}\\
\\
\indent \indent flm[[lvl]] <- u[[lvl]] * flm[[lvl]]\\
\\
\indent \indent cm <- cov.mat[[lvl + 1]] <- t(flm[[lvl]]) \%*\% cor.mat \%*\% flm[[lvl]]\\
\indent \indent cor.mat <- calc.cor.mat(cm)\\
\indent \}\\
\\
\indent if(!mkt.fac)\\
\indent \indent mod.mat <- cm\\
\indent else\\
\indent \{\\
\indent \indent k <- nrow(cor.mat)\\
\indent \indent x <- eigen(cor.mat)\\
\indent \indent y <- x\$vectors\\
\indent \indent y <- y[, 1]\\
\indent \indent z <- x\$values[1]\\
\indent \indent \\
\indent \indent mod.mat <- matrix(z, k, k)\\
\indent \indent mod.mat <- t(mod.mat * y) * y\\
\indent \indent diag(mod.mat) <- 1\\
\indent \indent mod.mat <- calc.cov.mat(mod.mat, sqrt(diag(cm)))\\
\indent \}\\
\\
\indent for(lvl in length(ind):1)\\
\indent \{\\
\indent \indent fac.cov <- mod.mat\\
\indent \indent mod.mat <- flm[[lvl]] \%*\% mod.mat \%*\% t(flm[[lvl]])\\
\indent \indent sv <- diag(1 - mod.mat)\\
\indent \indent diag(mod.mat) <- 1\\
\indent \indent tr <- sqrt(diag(cov.mat[[lvl]]))\\
\indent \indent mod.mat <-  calc.cov.mat(mod.mat, tr)\\
\indent \}\\
\\
\indent if(!rm.sing.tkr)\\
\indent \{\\
\indent \indent take <- colSums(ind[[1]]) == 1\\
\indent \indent x <- diag(fac.cov)\\
\indent \indent x[take] <- 0\\
\indent \indent diag(fac.cov) <- x\\
\indent \indent if(sum(take) > 1)\\
\indent \indent	\indent	take <- rowSums(ind[[1]][, take]) == 1\\
\indent \indent	else if (sum(take) == 1)\\
\indent \indent	\indent	take <- as.vector(ind[[1]][, take]) == 1\\
\indent \indent	else\\
\indent \indent	\indent	take <- rep(F, nrow(ind[[1]]))\\
\indent \indent sv[take] <- 1\\
\indent \}\\
\\
\indent spec.risk <- tr * sqrt(sv)\\
\indent fac.load <- tr * flm[[1]]\\
\\
\indent v <- flm[[1]] / sv\\
\indent d <- solve(fac.cov) + t(flm[[1]]) \%*\% v\\
\indent inv <- diag(1 / sv) -  v \%*\% solve(d) \%*\% t(v)\\
\indent inv <- calc.cov.mat(inv, 1 / tr)\\
\\
\indent result <- new.env()\\
\indent result\$spec.risk <- spec.risk\\
\indent result\$fac.load <- fac.load \\
\indent result\$fac.cov <- fac.cov\\
\indent result\$cov.mat <- mod.mat\\
\indent result\$inv.cov <- inv\\
\indent result <- as.list(result)\\
\indent return(result)\\
\}
}}

\section{R Code: Optimizer with Constraints \& Bounds}\label{app.C}

{}In this appendix we give the R source code for the optimization algorithm with linear homogeneous constraints and position bounds we use in Section \ref{sub.opt}. This code is similar to the code for the bounded regression algorithm discussed in detail in (Kakushadze, 2015b) with one important difference, so our discussion here will be brief. The entry function is {\tt{\small bopt.calc.opt()}}. The {\tt{\small args()}} of {\tt{\small bopt.calc.opt()}} are: {\tt{\small ret}}, which is the $N$-vector of stock returns (for a given date); {\tt{\small load}}, a matrix whose columns are the coefficients of the homogeneous constraints, so {\tt{\small dim(load)[1]}} is $N$ ({\em e.g.}, if the sole constraint is the dollar neutrality constraint, then {\tt{\small load}} is an $N\times 1$ matrix with unit elements); {\tt{\small inv.cov}}, which is the $N\times N$ inverse factor model covariance matrix $\Gamma^{-1}_{ij}$; {\tt{\small upper}}, which is the $N$-vector of the upper bounds $w_i^+$ on the weights $w_i$ (see below); {\tt{\small lower}}, which is the $N$-vector of the lower bounds $w_i^-$ on the weights $w_i$; and {\tt{\small prec}}, which is the desired precision with which the output weights $w_i$, the $N$-vector of which {\tt{\small bopt.calc.opt()}} returns, must satisfy the normalization condition $\sum_{i=1}^N |w_i| = 1$. Here the weights are defined as $w_i\equiv H_i/I$ (the dollar holdings over the total investment level). See (Kakushadze, 2015b) for more detail.\footnote{\, The analog of the line {\tt{\small y <- t(load[Jt, ]) \%*\% w.ret[Jt, ]}} below, reads {\tt{\small y <- t(w.load[Jt, ]) \%*\% ret[Jt, ]}} in (Kakushadze, 2015b), where the analog of {\tt{\small inv.cov}} is diagonal and both lines give the same result; however, for non-diagonal {\tt{\small inv.cov}} here they do not.}\\
\\
{\tt{\small
\noindent bopt.calc.opt <- function (ret, load, inv.cov, upper, lower, prec = 1e-5) \\
\{\\
\indent	x <- bopt.gen.lm(ret, load, inv.cov)\\
\indent	ret <- ret / sum(abs(x))\\
\\
\indent	repeat\{\\
\indent\indent		x <- bopt.opt(ret, load, inv.cov, upper, lower)\\
\indent\indent		if(abs(sum(abs(x)) - 1) < prec)\\
\indent\indent\indent			break\\
\\
\indent\indent		ret <- ret / sum(abs(x))\\
\indent	\}\\
\\
\indent	return(x)\\
\}\\
\\
bopt.gen.lm <- function (x, y, z)\\
\{\\
\indent	if(is.vector(z))\\
\indent\indent		z <- diag(z)\\
\indent	if(is.vector(y))\\
\indent\indent		y <- matrix(y, length(y), 1)\\
\indent	if(is.vector(x))\\
\indent\indent		x <- matrix(x, length(x), 1)\\
\indent	y1 <- z \%*\% y\\
\indent	res <- (z - y1 \%*\% solve(t(y) \%*\% y1) \%*\% t(y1)) \%*\% x \\
\indent	return(res)\\
\}\\
\\
bopt.opt <- function (ret, load, inv.cov, upper, lower, tol = 1e-6) \\
\{\\
\indent	calc.bounds <- function(z, x)\\
\indent	\{\\
\indent\indent		q <- x - z\\
\indent\indent		p <- rep(NA, length(x))\\
\indent\indent		pp <- pmin(x, upper)\\
\indent\indent		pm <- pmax(x, lower)\\
\indent\indent		p[q > 0] <- pp[q > 0]\\
\indent\indent		p[q < 0] <- pm[q < 0]\\
\indent\indent		t <- (p - z)/q\\
\indent\indent		t <- min(t, na.rm = T)\\
\indent\indent		z <- z + t * q\\
\indent\indent		return(z)\\
\indent	\}\\
\\
\indent	if(!is.matrix(load))\\
\indent\indent		load <- matrix(load, length(load), 1)\\
\\
\indent	n <- nrow(load)\\
\indent	k <- ncol(load)\\
\\
\indent	ret <- matrix(ret, n, 1)\\
\indent	upper <- matrix(upper, n, 1)\\
\indent	lower <- matrix(lower, n, 1)\\
\indent	z <- inv.cov\\
\indent	w.load <- z \%*\% load\\
\indent	w.ret <- z \%*\% ret\\
\\
\indent	J <- rep(T, n)\\
\indent	Jp <- rep(F, n)\\
\indent	Jm <- rep(F, n)\\
\indent	z <- rep(0, n)\\
\\
\indent	repeat\{\\
\indent\indent		Jt <- J \& !Jp \& !Jm\\
\indent\indent		y <- t(load[Jt, ]) \%*\% w.ret[Jt, ]\\
\indent\indent		if(sum(Jp) > 1)\\
\indent\indent\indent			y <- y + t(load[Jp, ]) \%*\% upper[Jp, ]\\
\indent\indent		else if(sum(Jp) == 1)\\
\indent\indent\indent			y <- y + upper[Jp, ] * matrix(load[Jp, ], k, 1)\\
\indent\indent		if(sum(Jm) > 1)\\
\indent\indent\indent		y <- y + t(load[Jm, ]) \%*\% lower[Jm, ]\\
\indent\indent		else if(sum(Jm) == 1)\\
\indent\indent\indent			y <- y + lower[Jm, ] * matrix(load[Jm, ], k, 1)\\
\indent\indent		if(k > 1)\\
\indent\indent\indent			take <- colSums(abs(load[Jt, ])) > 0\\
\indent\indent		else\\
\indent\indent\indent			take <- T\\
\indent\indent		Q <- t(load[Jt, take]) \%*\% w.load[Jt, take]\\
\indent\indent		Q <- solve(Q)\\
\indent\indent		v <- Q \%*\% y[take]\\
\\
\indent\indent		xJp <- Jp\\
\indent\indent		xJm <- Jm\\
\\	
\indent\indent		x <- w.ret - w.load[, take] \%*\% v\\
\indent\indent		x[Jp, ] <- upper[Jp, ]\\
\indent\indent		x[Jm, ] <- lower[Jm, ]\\
\\
\indent\indent		z <- calc.bounds(z, x)\\
\indent\indent		Jp <- abs(z - upper) < tol\\
\indent\indent		Jm <- abs(z - lower) < tol\\
\\		
\indent\indent		if(all(Jp == xJp) \& all(Jm == xJm))\\
\indent\indent\indent			break\\
\indent	\}\\
\\
\indent	return(z)\\		
\}
}}

\section{DISCLAIMERS}\label{app.D}

{}Wherever the context so requires, the masculine gender includes the feminine and/or neuter, and the singular form includes the plural and {\em vice versa}. The author of this paper (``Author") and his affiliates including without limitation Quantigic$^\circledR$ Solutions LLC (``Author's Affiliates" or ``his Affiliates") make no implied or express warranties or any other representations whatsoever, including without limitation implied warranties of merchantability and fitness for a particular purpose, in connection with or with regard to the content of this paper including without limitation any code or algorithms contained herein (``Content").

{}The reader may use the Content solely at his/her/its own risk and the reader shall have no claims whatsoever against the Author or his Affiliates and the Author and his Affiliates shall have no liability whatsoever to the reader or any third party whatsoever for any loss, expense, opportunity cost, damages or any other adverse effects whatsoever relating to or arising from the use of the Content by the reader including without any limitation whatsoever: any direct, indirect, incidental, special, consequential or any other damages incurred by the reader, however caused and under any theory of liability; any loss of profit (whether incurred directly or indirectly), any loss of goodwill or reputation, any loss of data suffered, cost of procurement of substitute goods or services, or any other tangible or intangible loss; any reliance placed by the reader on the completeness, accuracy or existence of the Content or any other effect of using the Content; and any and all other adversities or negative effects the reader might encounter in using the Content irrespective of whether the Author or his Affiliates is or are or should have been aware of such adversities or negative effects.

{}The R code included in Appendix \ref{app.A}, Appendix \ref{app.B} and Appendix \ref{app.C} hereof is part of the copyrighted R code of Quantigic$^\circledR$ Solutions LLC and is provided herein with the express permission of Quantigic$^\circledR$ Solutions LLC. The copyright owner retains all rights, title and interest in and to its copyrighted source code included in Appendix \ref{app.A}, Appendix \ref{app.B} and Appendix \ref{app.C} hereof and any and all copyrights therefor.



\begin{table}[ht]
\noindent
\caption{First column: the number of principal components $K$; last column: $g(K)$ defined in (\ref{g}); Min, 1st Quartile, Median, Mean, 3rd Quartile and Max refer to the corresponding quantities for the ratio ${\widetilde \xi}_i^2 = \xi^2_i/C_{ii}$ (specific variance over total variance). The number of observations (days) in the time series is $M+1 = 20$. The number of (randomly selected) stocks is $N=2316$. All quantities are rounded to 3 digits. The value of $K$ fixed via (\ref{K}) is $K=12$. See Figure 1 for a density plot.} 
\begin{tabular}{l l l l l l l l} 
\\
\hline\hline 
$K$ & Min & 1st Quartile & Median & Mean & 3rd Quartile & Max & $g(K)$ \\[0.5ex] 
\hline 
1 & 0.16 & 0.62 & 0.824 & 0.771 & 0.953 & 1 & 0.4 \\
2 & 0.137 & 0.525 & 0.693 & 0.682 & 0.867 & 1 & 0.37 \\
3 & 0.084 & 0.453 & 0.629 & 0.618 & 0.8 & 0.999 & 0.29 \\
4 & 0.075 & 0.405 & 0.562 & 0.56 & 0.718 & 0.992 & 0.27 \\
5 & 0.06 & 0.355 & 0.501 & 0.51 & 0.668 & 0.981 & 0.235 \\
6 & 0.06 & 0.312 & 0.449 & 0.462 & 0.606 & 0.977 & 0.233 \\
7 & 0.057 & 0.272 & 0.396 & 0.417 & 0.552 & 0.931 & 0.203 \\
8 & 0.033 & 0.233 & 0.347 & 0.375 & 0.503 & 0.916 & 0.139 \\
9 & 0.029 & 0.203 & 0.306 & 0.334 & 0.446 & 0.884 & 0.111 \\
10 & 0.019 & 0.172 & 0.264 & 0.294 & 0.39 & 0.84 & 0.056 \\
11 & 0.01 & 0.144 & 0.227 & 0.256 & 0.339 & 0.84 & 0.018 \\
12 & 0.009 & 0.118 & 0.194 & 0.22 & 0.294 & 0.84 & 0.013 \\
13 & 0.008 & 0.093 & 0.157 & 0.186 & 0.25 & 0.728 & 0.056 \\
14 & 0.002 & 0.07 & 0.122 & 0.152 & 0.204 & 0.696 & 0.12 \\
15 & 0.002 & 0.05 & 0.089 & 0.119 & 0.162 & 0.686 & 0.128 \\
16 & 0 & 0.029 & 0.062 & 0.088 & 0.122 & 0.608 & 0.211 \\
17 & 0 & 0.014 & 0.035 & 0.057 & 0.077 & 0.606 & 0.221 \\
18 & 0 & 0.003 & 0.011 & 0.028 & 0.034 & 0.592 & 0.231 \\ [1ex] 
\hline 
\end{tabular}
\label{table.prin.comp} 
\end{table}

\begin{table}[ht]
\caption{Simulation results for the weighted regression alphas discussed in Section \ref{sub.reg} and the optimized alphas discussed in Section \ref{sub.opt}, without any bounds on the dollar holdings. All quantities are rounded to 2 digits. See Figure 2 for P\&L plots.} 
\begin{tabular}{l l l l} 
\\
\hline\hline 
Alpha & ROC & SR & CPS\\[0.5ex] 
\hline 
Regression: Principal Components & 46.80\% & 11.50 & 2.05\\
Optimization: Principal Components & 47.74\% & 11.88 & 2.26\\
Regression: BICS Sub-Industries & 49.36\% & 12.89 & 2.16\\
Regression: Heterotic Risk Factors & 51.89\% & 13.63 & 2.27\\
Optimization: Heterotic Risk Model & 55.90\% & 15.41 & 2.67\\[1ex] 

\hline 
\end{tabular}
\label{table2} 
\end{table}

\begin{table}[ht]
\caption{Simulation results for the weighted regression alphas discussed in Section \ref{sub.reg} and the optimized alphas discussed in Section \ref{sub.opt}, with the liquidity bounds (\ref{liq}) on the dollar holdings. All quantities are rounded to 2 digits. See Figure 3 for P\&L plots.} 
\begin{tabular}{l l l l} 
\\
\hline\hline 
Alpha & ROC & SR & CPS\\[0.5ex] 
\hline 
Regression: Principal Components & 41.27\% & 14.24 & 1.84\\
Optimization: Principal Components & 40.92\% & 14.33 & 1.96\\
Regression: BICS Sub-Industries & 44.56\% & 16.51 & 1.97\\
Regression: Heterotic Risk Factors & 46.86\% & 18.30 & 2.08\\
Optimization: Heterotic Risk Model & 49.00\% & 19.23 & 2.36\\[1ex] 
\hline 
\end{tabular}
\label{table3} 
\end{table}

\newpage
\begin{figure}[ht]
\centerline{\epsfxsize 4.truein \epsfysize 4.truein\epsfbox{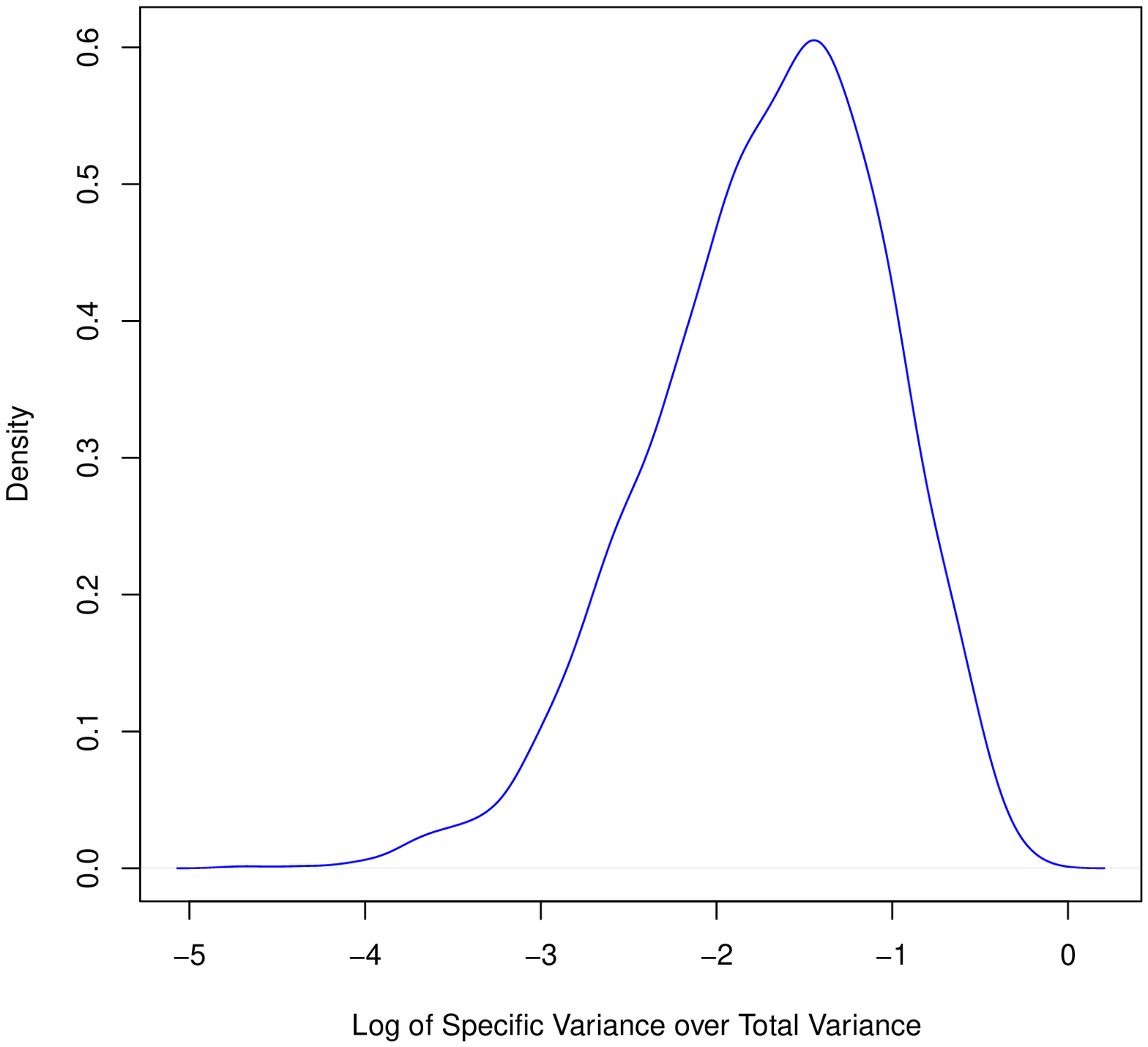}}
\noindent{\small {Figure 1. The density (computed using the R function {\tt{\small{density()}}}) for the log of the ratio ${\widetilde\xi}_i^2 = \xi^2_i/C_{ii}$ (specific variance over total variance) for the $K=12$ case in Table \ref{table.prin.comp}.}}
\end{figure}

\begin{figure}[ht]
\centerline{\epsfxsize 4.truein \epsfysize 4.truein\epsfbox{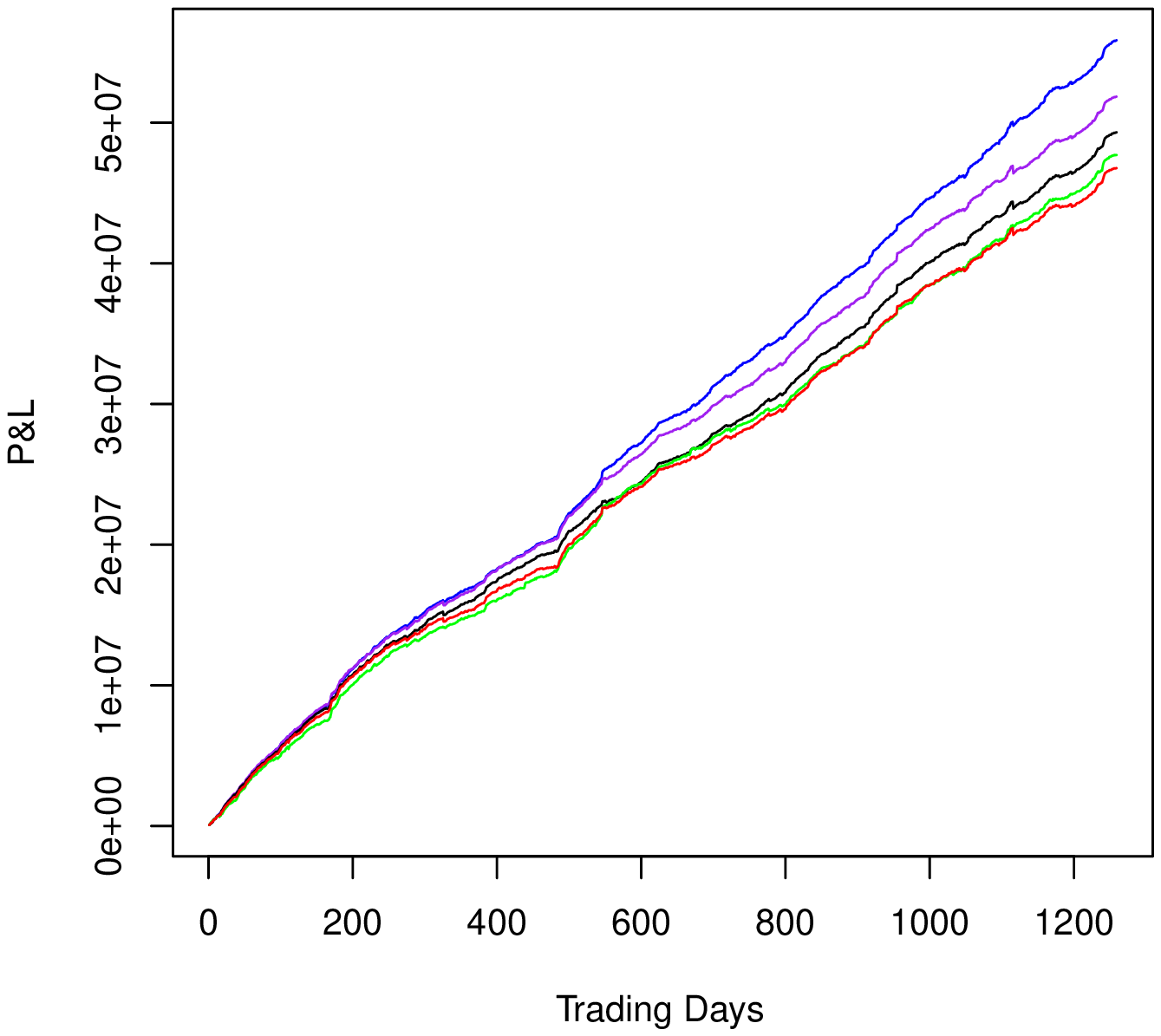}}
\noindent{\small {Figure 2. P\&L graphs for the intraday alphas without liquidity bounds summarized in Table \ref{table2}. Bottom-to-top-performing: i) regression over 20 principal components (Section \ref{sub.reg}), ii) optimization using the principal component risk model (Section \ref{sub.opt}), iii) regression over the BICS sub-industries (Section \ref{sub.reg}), iv) regression over the heterotic risk factor loadings (Section \ref{sub.reg}), and v) optimization using the heterotic risk model (Section \ref{sub.opt}). The investment level is \$10M long plus \$10M short.}}
\end{figure}

\begin{figure}[ht]
\centerline{\epsfxsize 4.truein \epsfysize 4.truein\epsfbox{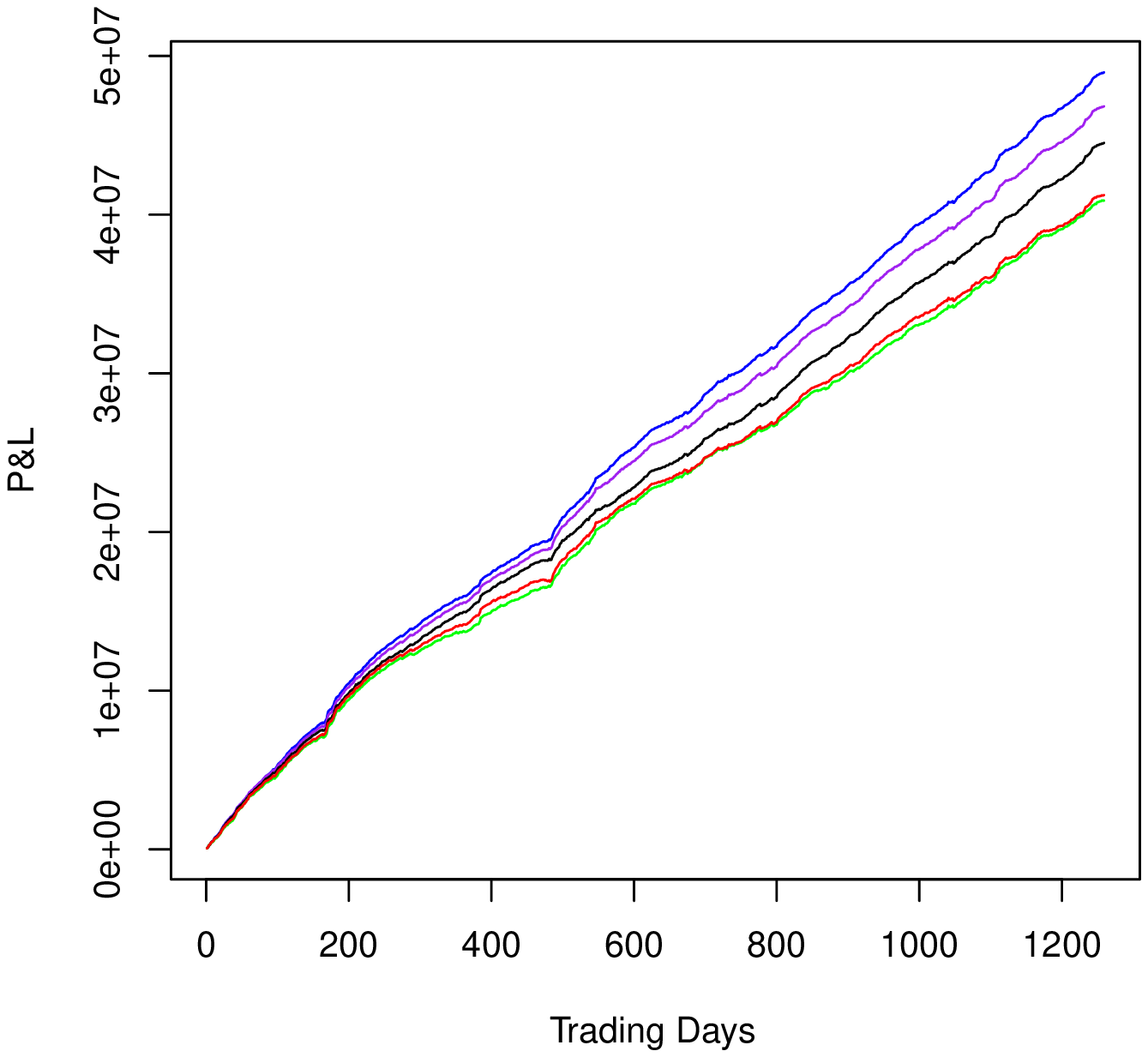}}
\noindent{\small {Figure 3. P\&L graphs for the intraday alphas with liquidity bounds summarized in Table \ref{table3}. Bottom-to-top-performing: i) optimization using the principal component risk model (Section \ref{sub.opt}), ii) regression over 20 principal components (Section \ref{sub.reg}), iii) regression over the BICS sub-industries (Section \ref{sub.reg}), iv) regression over the heterotic risk factor loadings (Section \ref{sub.reg}), and v) optimization using the heterotic risk model (Section \ref{sub.opt}). The investment level is \$10M long plus \$10M short.}}
\end{figure}

\end{document}